\documentclass[fleqn,usenatbib]{mnras}

\usepackage{newtxtext,newtxmath}

\usepackage[T1]{fontenc}
\usepackage{ae,aecompl}

\usepackage{graphicx}	
\usepackage{mathtools}
\usepackage{amsmath}	
\usepackage[utf8]{inputenc}
\usepackage{xcolor}    
\usepackage{multirow}  
\usepackage{epsfig}
\usepackage{nicefrac}
\usepackage{booktabs}   
\usepackage{ulem}
\usepackage{array}      
\usepackage{siunitx}    
\usepackage{orcidlink}
\usepackage{threeparttable}

\newcommand\catname{SDSS~J020536-081424\,}
    
\newcommand\RefStar{Gaia~EDR3~2463146603042188160\,}

\newcommand\Ks{K$_{s}$\,}

\title[Investigating the clumpy SF in \catname]{Investigating the clumpy star formation in an interacting dwarf irregular galaxy}

\author[Lassen et al.]{Augusto E. Lassen\textsuperscript{
\orcidlink{0000-0003-3575-8316}\,1}\thanks{augusto.lassen@gmail.com}, 
Ana L. Chies-Santos\textsuperscript{
\orcidlink{0000-0003-3220-0165}\,1},
Rogerio Riffel\textsuperscript{
\orcidlink{0000-0002-1321-1320}\,1},
Evelyn J. Johnston\textsuperscript{
\orcidlink{0000-0002-2368-6469}\,2},\newauthor
Eleazar R. Carrasco\textsuperscript{
\orcidlink{0000-0002-7272-9234}\,3},
Boris H\"au\ss ler\textsuperscript{
\orcidlink{0000-0002-1857-2088}\,4},
Gabriel M. Azevedo\textsuperscript{
\orcidlink{0000-0002-5445-3326}\,1,5},
Jean M. Gomes\textsuperscript{
\orcidlink{0000-0002-3861-7482}\,6},\newauthor
Rogemar A. Riffel\textsuperscript{
\orcidlink{0000-0003-0483-3723}\,7},
Ariel Werle\textsuperscript{
\orcidlink{0000-0002-4382-8081}\,8},
Rubens E. G. Machado\textsuperscript{
\orcidlink{0000-0001-7319-297X}\,9},
Daniel Ruschel-Dutra\textsuperscript{
\orcidlink{0000-0002-2690-9728}\,10}
\vspace{0.3cm}
\\
\textsuperscript{1}Instituto de Física, Departamento de Astronomia, Universidade Federal do Rio Grande do Sul, Av. Bento Gonçalves 9500, Porto Alegre, RS, Brazil\\
\textsuperscript{2}Instituto de Estudios Astrofísicos, Facultad de Ingeniería y Ciencias, Universidad Diego Portales, Av. Ejército Libertador 441, Santiago, Chile\\
\textsuperscript{3} International Gemini Observatory/NSF NOIRLab, Casilla 603, La Serena, Chile\\
\textsuperscript{4} European Southern Observatory, Alonso de Cordova 3107, Casilla 19001, Santiago, Chile\\
\textsuperscript{5} Departamento de Física Teórica y del Cosmos, Facultad de Ciencias, Universidad de Granada, Av. de Fuente Nueva, Granada, Spain\\
\textsuperscript{6} DTx–Digital Transformation CoLab, Building 1, Azurém Campus, University of Minho, PT4800-058 Guimarães, Portugal\\
\textsuperscript{7} Departamento de Física, Centro de Ciências Naturais e Exatas, Universidade Federal de Santa Maria, 97105-900, Santa Maria, RS, Brazil\\
\textsuperscript{8} INAF – Osservatorio Astronomico di Padova, Vicolo dell’Osservatorio 5, 35122 Padova, Italy\\
\textsuperscript{9} Departamento Acad\^emico de F\'isica, Universidade Tecnol\'ogica Federal do Paran\'a, Av. Sete de Setembro 3165, Curitiba, PR, Brazil
\\
\textsuperscript{10} Departamento de F\'isica, Universidade Federal de Santa Catarina, P.O. Box 476, 88040-900 Florian\'opolis, SC, Brazil
}

\date{Accepted XXX. Received YYY; in original form ZZZ}

\pubyear{2025}

\begin{document}
\label{firstpage}
\pagerange{\pageref{firstpage}--\pageref{lastpage}}
\maketitle

\begin{abstract}
Clumpy morphologies are more frequent in distant and low-mass star-forming galaxies. Therefore the less numerous nearby galaxies presenting kpc-sized clumps represent unique laboratories from which to address the mechanisms driving clump formation and study why such structures become less common in the local Universe, and why they tend to exhibit smaller sizes and lower star formation rates compared to their high-$z$ counterparts. We use high spatial resolution Integral Field Unit observations from VLT/MUSE to investigate the properties of several kpc-sized clumps seen in \catname, a $z \approx 0.04$ dwarf irregular galaxy interacting with its more massive companion Mrk~1172 ($\log (M/M_{\odot}) \sim 11$).
H$\alpha$ channel maps reveal that the clumps are embedded within a rotating ionised gas component, possibly a disk. Self-consistent full-spectral fitting of the clump spectra with \textsc{fado} indicates that their young ($t \leq 10$\,Myr) populations have lower stellar metallicities compared to the older ($t \gtrsim 100$\,Myr) ones, although these estimates are subject to significant degeneracies. The clumpy SF in \catname seems to occur in the disk, which dominates the stellar emission. Gas-phase metallicities derived through strong-line calibrations exhibit a flat distribution around $Z_{\mathrm{gas}} \approx 0.3\,Z_{\odot}$, likely caused by efficient galactic-scale metal mixing processes. There is no evidence for a strong anti-correlation between $Z_{\mathrm{gas}}$ and SFR, although clump sizes are likely overestimated due to seeing limitations. 
The lower $Z_{\ast}$ of younger stellar populations compared to the disk suggests clump formation driven by accretion of metal-poor gas in \catname.
\end{abstract}


\section{Introduction}
\label{sec:intro}
The morphologies of distant ($z \gtrsim 1$) galaxies tend to be more asymmetric than that of nearby galaxies: Recent observations with the James Webb Space Telescope (JWST) indicate that the fraction of prolate galaxies among low-mass ($9.0 < \log \,(M_{\star }/M_{\odot}) < 9.5$) galaxies increases with redshift, meanwhile the fraction of disks in dwarfs increases towards the local Universe \citep{Pandya2024}. The morphology of $z \gtrsim 1$ star-forming galaxies is also more irregular in comparison to local galaxies \citep{Conselice2000,Elmegreen2005a,Fisher2017a, Huertas-Company2023}, being dominated by massive kpc-sized ``clumps'' \citep{Cowie1995,Bergh1996,Elmegreen2005b,Dekel2009}. The fraction of galaxies with Ultra-Violet (UV) clumps was observed to increase with redshift in a trend that roughly follows the evolution of the Star Formation Rate Density \citep[SFRD,][]{Madau2014, Shibuya2016}. 

In addition to the redshift, the fraction of clumpy Star-Forming Galaxies (SFGs) also depends on the stellar mass: Considering the range $0.5 < z < 3$, up to 60 per cent of low-mass galaxies ($M_{\star} \lesssim 10^{9.8}\,M_{\odot}$) present clumpy UV morphologies, while this fraction drops to 15--55 per cent in the case of high-mass SFGs in the same redshift range \citep{Guo2018,Forster-Schreiber2020}. With overall contributions to the galactic rest-frame UV emission in the range of 20--40 per cent, the clumps correspond to sites of elevated Star Formation Rate (SFR), reaching up to 10\,$M_{\odot}$\,yr$^{-1}$ per clump \citep{Elmegreen2005b,Genzel2011,Soto2017,Fisher2017a,Forster-Schreiber2020}. The stellar mass of typical individual clumps lie in the range of $10^{8}\,-\,10^{9}\,M_{\odot}$ \citep{Wisnioski2012,Tacconi2013,Messa2019}, thereby the clumps are not only larger, brighter and more massive than local star-forming regions, but also presents much higher SFRs and specific SFRs (sSFR) \citep{Guo2012,Fisher2017b}. Observations of clumps in lensed galaxies and high-resolution simulations suggest, however, that clump properties such as sizes and stellar masses are systematically overestimated due to resolution limitations, e.g., if the physical size of the clumps is smaller than the imaging resolution, several clumps will be blended as a source artificially boosted towards high stellar masses \citep{Elmegreen2013,Dessauges2017, Cava2018}.

In the last decades there have been substantial efforts to explain how these clumps formed and why they are so much rarer in Local Universe galaxies, giving rise to two major scenarios. The first scenario asserts that clumps are formed from major or minor mergers during the early build-up of the host galaxy, when the satellite is stripped and its nucleus becomes the massive \textit{ex situ} clump or generates local instabilities within the host galaxy that drive further collapse of the gas \citep{Somerville2001,Hopkins2013,Ribeiro2017}. Observations and simulations suggest that only a minority of the high-$z$ clumps are formed this way and that clump formation depends on the galaxy mass, so that the \textit{ex situ} mechanism is more common for $M \lesssim 10^{10}\,M_{\odot}$ galaxies \citep{Mandelker2014,Guo2015,Zanella2019,Adams2022}. The second and most supported scenario is the \textit{in situ} formation of the clumps, generated from the collapse of gas-rich disk fragments due to Violent Disk Instability \citep[VDI,][]{Noguchi1999,Dekel2009,Agertz2009,Bournaud2014}. Smooth cold accretion of intergalactic gas, for instance, can drive disks to be marginally stable \citep[i.e., the Toomre parameter is $Q \sim$1;][]{Toomre1964}, causing the gravitational instabilities that give rise to the massive clumps \citep{Genzel2008,Bouche2013,Adams2022}. At high-$z$, disks can fragment at much larger scales than their local-Universe counterparts, given that they are highly turbulent \citep[$\sigma_{\mathrm{H}\alpha}\sim 30 - 80\,$km\,s$^{-1}$,][]{Green2010,Bassett2014} and present gas-to-stellar mass ratios two times larger than local disks \citep{Tacconi2008,Tacconi2013,Saintonge2013}, a combination of ingredients that increase the Jeans lengths at the same time that large gas reservoirs fuel the star formation. The \textit{in situ} interpretation is supported by observations in $z \sim$ 1 galaxies \citep{Zanella2015,Dessauges2019} and local analogues \citep{Fisher2017a,Fisher2017b,Messa2019} and by simulations of turbulent high-$z$ galaxies \citep{Donkelaar2022}. In accordance with that, simulations predict that galaxies with baryonic mass fractions $\gtrsim 0.3$ have turbulent Interstellar Medium (ISM) and clumpy morphology \citep{Hayward2017}. In the Local Universe such turbulent ISM is found almost exclusively in low-mass systems \citep{Bradford2015}. 

In the the hierarchical framework of galaxy formation, metal-poor dwarf galaxies are considered the building blocks of the present-day larger galaxies \citep{White91,DeLucia2007,Annibali2022}. Furthermore, star formation follows a hierarchical behaviour \citep{Lada2003}, so that the clumpy morphologies frequently observed in high-$z$ galaxies are associated with early stages of star formation. Notwithstanding, the fraction of clumpy galaxies is uncommon among local galaxies \citep[$z \sim 0$; e.g., ][]{Shibuya2016,Fisher2017a}, there are few local metal-poor starburst dwarf galaxies that represent the most promising analogues for typical high-$z$ clumpy galaxies. Besides the metal-poor content, they usually present strong stellar feedback and high \ion{H}{i} mass fractions \citep{Muratov2015,Trebitsch2017}.
The large distances to high-$z$ galaxies do not allow us to spatially resolve the sub-structures that make up the clumps, so these local analogues enable us to explore the physics underlying star formation processes at intermediate scales 
\citep[i.e., at higher scales than the typical $\sim$10\,pc of Young Star Clusters and smaller than the kpc-sized giant stellar clumps;][]{Messa2019, Adamo2020} and to constrain stellar processes in metal-deficient environments.

\begin{figure*}
    \centering
    \includegraphics{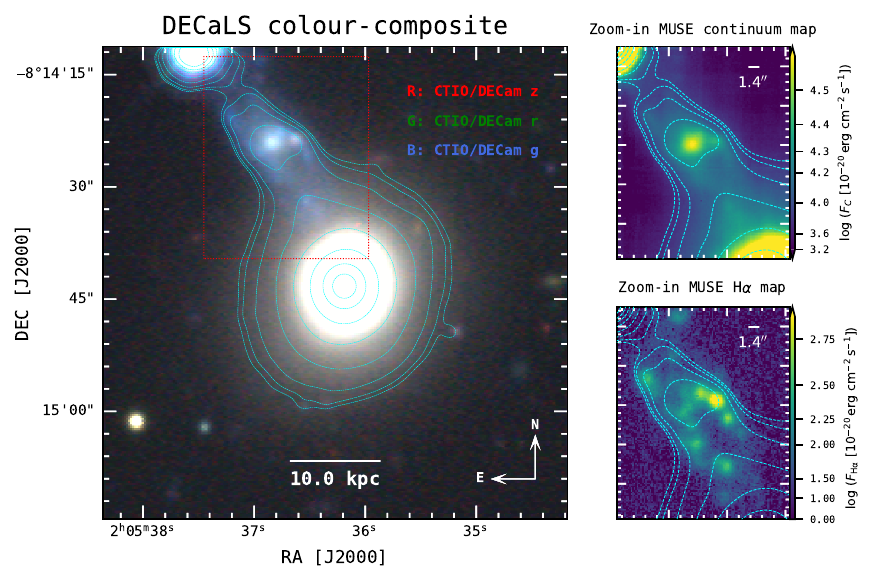}
    \caption{
    \textit{Left:} DECaLS colour-composite image of the dwarf-ETG system analysed in this work, cropped to match the 1 arcmin$^{2}$ FoV of the VLT/MUSE observations. The ETG Mrk~1172 is located at the centre of the image, and the bright source on the top left corner is a catalogued star. Dotted cyan contours represent isophote levels ranging from $\sim$18\,mag\,arcsec$^{-2}$ to $\sim$21.5\,mag\,arcsec$^{-2}$ in CTIO/DECam-$z$ band. The orange dotted rectangle encompasses the dwarf galaxy \catname and indicates the zoomed-in region shown in the right-hand panels. The images were smoothed with \textsc{inla} and coloured with \textsc{trilogy}, with the single exposures used as RGB channels indicated in the top right.
    \textit{Top Right:} Zoom-in MUSE continuum map of the dwarf. The scale bar represents the approximate PSF size ($\sim$1.1\,kpc).
    \textit{Bottom Right:} Same region of top right panel, but now in the continuum-subtracted H$\alpha$, showing the presence of several off-centre star-forming clumps. The contours shown in the right-hand images are the same displayed in the DECaLS image.}
    \label{fig:MUSE_FoV}
\end{figure*}

In \cite{Lassen2021} (L21 hereafter) we studied the local ($z = 0.04025$) dwarf irregular companion of the Early-Type Galaxy (ETG) Mrk~1172, \catname. The dwarf galaxy occupies a projected region of 14$\times$ 14\,kpc$^{2}$ ($\sim$10\arcsec) and, although faint in the stellar continuum, is bright in H$\alpha$ and presents a clumpy morphology (see Fig.~\ref{fig:MUSE_FoV}). 
The typical radii of the clumps are of the order of the observational seeing ($\sim$1.4\arcsec), which corresponds to a Point Spread Function (PSF) radius of $\sim$1 kpc. At this spatial scale, PSF-convolved star-forming complexes likely represent a blend of smaller star-forming regions with surrounding ionised gas. \catname represents an interesting system to study: It lies at a considerably large distance ($d_L \sim $185\,Mpc) compared to typical Integral Field Unit (IFU) studies of low-mass galaxies \citep[e.g.][among many others]{James2020, Herenz2023, Macarena2023} but at the same time it is close enough to identify and analyse the clumps separately, correlating with properties from the host low-mass galaxy. Furthermore in L21 it was shown that \catname is in interaction with the Early-Type Galaxy (ETG) Mrk~1172, turning this system into a unique opportunity to use the powerful spatially resolved analysis from IFU spectroscopy to explore scenarios of clump formation. By comparing the metallicities of clumps and underlying ionised gas component, we also aim to investigate the scenario of infall of metal-poor gas as the trigger of clumpy star formation.

Throughout this work we adopt the $\Lambda$CDM cosmological parameters with a Hubble constant of $H_{0} = 67.4\,\rm{km}\,\rm{s}^{-1}\,\rm{Mpc}^{-1}$ and density parameters ($\Omega_{\rm m}$, $\Omega_{\Lambda}$) = 0.315, 0.685, as reported by \cite{Planck2020} results, and Oxygen abundance as a tracer of the overall gas phase metallicity, using the two terms interchangeably, adopting $\rm{log(O/H)} + 12 = 8.69$ as the solar Oxygen abundance \citep{Asplund2021}. In Sect.~\ref{sec:MUSE_data} we outline the data and summarise key findings previously determined in L21. 
In Sect.~\ref{sec:sec3} we detail the methodologies employed to extract and analyse clump spectra. In Sect.~\ref{sec:results} we show the main findings of this work, which are discussed in Sect.~\ref{sec:discussion}. We summarise our main results and and outline the conclusions in Sect.~\ref{sec:conclusion}.

\section{Data description}
\label{sec:MUSE_data}
The dwarf-ETG system was observed with the integral field spectroscopy Multi-Unit Spectroscopic Explorer \citep[MUSE,][]{Bacon2010} mounted at the $8.2\,$m ESO Very Large Telescope (VLT) on Paranal (Chile), using the Wide-Field Mode (WFM; P.I.: E.J. Johnston). The WFM covers the nominal wavelength range of $4600\,$--$9350\,$\AA\, with mean spectral sampling of $1.25\,$\AA\, over a 1~arcmin$^{2}$ field-of-view with angular sampling of 0.2\arcsec. The resolving power ($R \equiv \Delta\lambda/\lambda$) of the MUSE spectrograph varies from $R \sim$1750 at 465\,nm to $R \sim$3750 at 930\,nm. In L21, a spatially resolved characterisation of several physical and chemical properties of the dwarf galaxy was presented, which we briefly summarise here. For details on the data reduction process, we refer to L21.

The integrated instantaneous SFR (i.e., the Star Formation Rate in timescales of $\sim$10\,Myr), estimated using the \cite{Kennicutt1998} calibration with L(H$\alpha$), resulted in 0.7\,$M_{\odot}\,\mathrm{yr}^{-1}$, assuming a Salpeter stellar Initial Mass Function (IMF). Using stellar population synthesis we estimated a total stellar mass of $3 \times 10^{9}\,M_{\odot}$, hence leading to $\mathrm{log (sSFR)} = -9.63$. The mean gas-phase metallicity, estimated from the strong-line calibrations of \cite{Marino2013} and \cite{Dopita2016}, was 12 + $\log$(O/H) = $8.2 \pm 0.2$, indicating a metal-deficient ($\sim$\nicefrac{1}{3}\,$Z_{\odot}$) ISM. A kinematic model assuming gas in circular orbits was fitted to the system using the stellar and gas velocity field of the ETG and the dwarf, respectively. The alignment of both line of the nodes and disk inclination indicates that the gas in the dwarf is under the gravitational influence of Mrk~1172.

We present the dwarf-ETG system in Fig.~\ref{fig:MUSE_FoV}, where the left panel shows a colour-composite image from the Dark Energy Camera Legacy Survey \citep[DECaLS,][]{Dey2019} cropped within the 1\,arcmin$^{2}$ FoV observed by MUSE. The image was coloured using \textsc{trilogy} \citep{Coe2012}, by combining exposures in the $z$, $r$, and $g$ filters as RGB channels, respectively. The ETG is located at the centre of the image, while the dotted orange rectangle encompasses the dwarf galaxy and indicates the zoomed-in region shown in the right-hand panels. The bright source at the top of the image is the star \RefStar. The dotted cyan contours correspond to the isophote levels of $17.9 \lesssim \mu_{\mathrm{AB}, z}\,[\mathrm{mag/arcsec}^{2}] \lesssim 21.5$. In the upper and lower right panels, we display zoom-in emission maps of the dwarf in the rest-frame continuum window from 4570 \AA\, to 7000 \AA\,(considering the dwarf reference frame) and in the continuum-subtracted H$\alpha$, respectively. For the latter, the continuum modelling was achieved by applying a median filter with a spectral width of 150~\AA\,\citep{Herenz2017}. 

The images displayed in Fig.~\ref{fig:MUSE_FoV} were smoothed using the Integrated Nested Laplace Approximation method
\citep[\textsc{inla\footnote{\label{foot:INLA_link}\url{https://www.r-inla.org/home}}},][]{Rue2009}
to account for the spatial correlations between adjacent spaxels, with the corresponding variance maps employed to build the weighted covariance matrices. For previous applications of \textsc{inla} to smooth astronomical
spatially resolved maps, we refer to the works of \cite{Gonzalez-Gaitan2019} and \cite{Azevedo2023}. From the bottom right panel, it is possible to observe that the morphology of \catname in H$\alpha$ is dominated by off-centre star-forming clumps. In Appendix~\ref{appendix:GSAOI} we also present observations of \catname in the Near-Infrared (NIR) using the Gemini South Adaptive Optics Imager \citep[GSAOI,][]{Mcgregor2004, Carrasco2012}. The high spatial resolution of GSAOI provides images with enhanced seeing, but only the brightest clump could be detected in the NIR, preventing to dig deeper on the internal structure of \catname and quantify by how much the clump sizes are overestimated from IFU-based estimates.

\section{Methodology}
\label{sec:sec3}

Before applying the spectroscopic methods presented in this chapter, the original datacube was shifted to the rest frame using the $z = 0.04025$ estimate from L21 and corrected for Milky Way foreground extinction with the dust maps from \cite{Schlafly2011}, assuming a homogeneous dust screen model described by the CCM reddening law \citep{cardelli1989}, with $R_{V} = 3.1$ and $(\mathrm{H}\alpha / \mathrm{H}\beta)_{\mathrm{int}} = 2.863$ \citep{Osterbrock2006}.

\subsection{Detection and extraction of individual clump spectra}
\label{sec:clump_regions}
As observed in Fig.~\ref{fig:MUSE_FoV}, the 2D H$\alpha$ morphological map of \catname is dominated by several star-forming clumps superimposed on an extended component of ionised gas. To conduct a spectroscopic analysis of the clumps, it is essential to establish clear criteria to separate the compact structures from the adjacent extended gas component.

The identification of the clumps was performed using \textsc{astrodendro}\footnote{\label{foot:astrodendro_link}\url{https://dendrograms.readthedocs.io/en/stable/}}
on the continuum-subtracted H$\alpha$ image to build a hierarchical tree-diagram of structures, which are categorised in branches (i.e. structures that split into multiple sub-structures, including other branches), and leaves, that do not contain sub-structures. The construction of a dendrogram is based on an isophotal-like analysis, since the code starts from the highest-flux pixel of the image and assigns subsequent pixels to branches and leaves based on the flux intensity level \citep{Goodman2009}. We allow detection from a minimum flux threshold of $F(\rm{H}\alpha) = 5.2 \times 10^{-19}\,\rm{erg}\,\rm{s}^{-1}\,\rm{cm}^{-2}$, corresponding to approximately 5\,$\times$ the background RMS estimated from an empty region on the H$\alpha$ image, and for a minimum size of 15 pixels per source ($\approx 0.4\,\rm{kpc}^{2}$). For the \verb|min_delta| parameter we use the same flux value, meaning that a peak must have a flux at least 5$\times$ the background RMS to be considered as a new independent structure.

\begin{figure}
    \centering
    \includegraphics[scale=0.85]{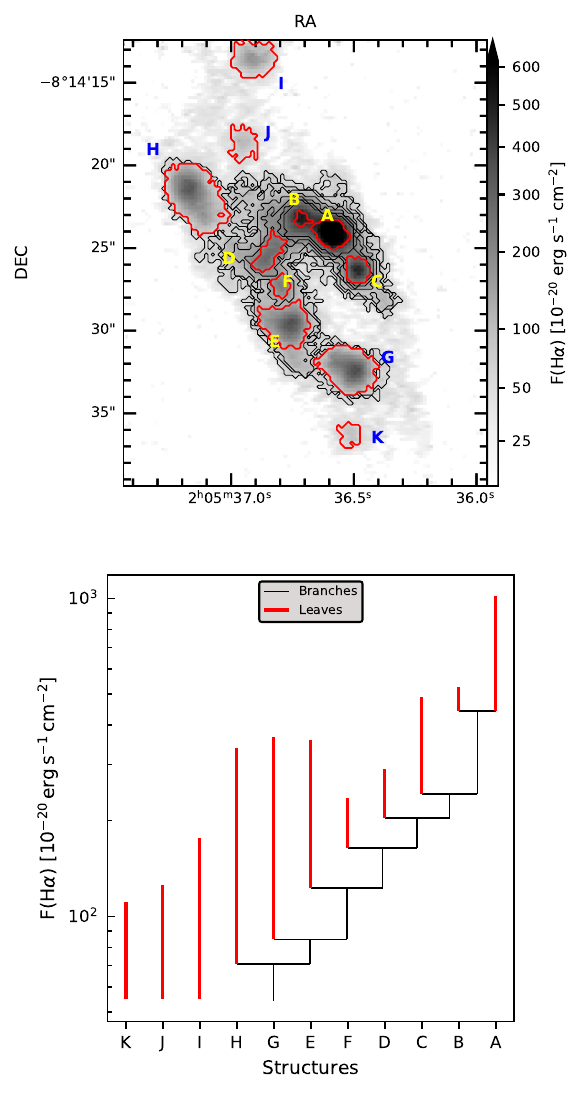}
    \caption{\textit{Top panel}: Continuum-subtracted H$\alpha$ image of \catname with overlaid hierarchical contours from \textsc{astrodendro}. Black and red contours stand for structural branches and leaves (i.e. structures with no sub-structures), respectively.
    In yellow and blue we present the labels assigned to each leaf, according to their F(H$\alpha$) surface density. The map was smoothed using \textsc{inla} method.
    \textit{Bottom panel}: Hierarchical tree-diagram of the detected structures. The flux scale on the $y$ axis shows the integrated flux of the structures, which were labelled according to their F(H$\alpha$) surface density.}
    \label{fig:knots}
\end{figure}

The hierarchical flux levels are presented in the top panel of Fig.~\ref{fig:knots}. The black contours represent the flux level of the branches, highlighting the separation between the ionised gas component from the background level. The red contours represent the leaves, which include clumps separated from the extended component and islands of flux (Clumps I, J and K). In the bottom panel, we present the resulting hierarchical tree-diagram, where the logarithmic flux scale stands for the integrated F(H$\alpha$) of each structure. Since the area covered by each clump can vary significantly, we label them in alphabetic order according to their flux surface density, a better tracer to rank compact star-formation structures than the integrated flux alone. From the detected structures in Fig.~\ref{fig:knots} we extract the integrated spectrum of each clump using a PSF-sized aperture centred at the emission peak of the corresponding leaf.

\subsection{Self-consistent continuum modelling}
\label{subsec:continuum}
To accurately assess the intensity of nebular lines in the clump spectra, it is essential to address the continuum (stellar plus nebular) appropriately, especially in highly ionised regions. It is crucial to account for the absorption features of the underlying stellar populations, especially when measuring the profiles and intensities of emission lines like H$\alpha$ and H$\beta$. This correction becomes significant when the contributions from young and intermediate stellar populations are non-negligible \citep{Gonzalez1999a,Gonzalez1999b}. A common approach to model the stellar emission is via stellar population synthesis techniques, under the assumption that the stellar emission can be fairly described by linear combinations of Simple Stellar Populations (SSPs), e.g. 
\textsc{pPXF} \citep{Cappellari2004, Cappellari2017}, 
\textsc{starlight} \citep{starlight2005, Werle2019}, 
\textsc{fado} \citep{Gomes2017}, 
\textsc{pipe3D} \citep{Sanchez2016, Lacerda2022}, among many others.

In the absence of a massive underlying old population, nebular emission can be significant in regions of high star formation activity, reaching contributions to the total optical/IR continuum emission of up to 70\% \citep{Krueger1995,Gomes2017,Cardoso2019}. The nebular emission can be divided into two components: The first component stands for the forbidden and fine structure line transitions caused by the ionising radiation field produced by the short-lived massive stars, and the second component stands for the nebular continuum originated due to free-free, free-bound and two-photon emission \citep{Molla2009,Reines2010, Byler2017}. When the contribution from the nebular component becomes significant, stellar population fitting codes were reported to overestimate both the fraction of older stellar populations and the stellar mass \citep{Izotov2011}.

With this in mind, we perform a stellar population synthesis with \textsc{fado}
\citep[\textbf{F}itting \textbf{A}nalysis using \textbf{D}ifferential evolution \textbf{O}ptimisation,][]{Gomes2017} to model the continuum of the extracted clump spectra self-consistently, i.e. in addition to the stellar emission, flux estimates of the strongest nebular lines in the spectra are used to infer the nebular continuum and the number of young stars necessary to achieve the estimated ionisation parameter $U$. In this work we adopted the SSP models from \cite{Bruzual2003}\footnote{The incorporation of BC03 SSP models into \textsc{fado} is crucial for encompassing  the UV-ionising spectral range emitted by stellar populations younger than $20\,$Myr. These are responsible in ionising the majority of the interstellar gas within star-forming regions.} with the PADOVA\,1994 \citep{Alongi1993, Bressan1993, Fagotto1994a, Fagotto1994b, Girardi1996} evolutionary tracks, STELIB stellar library \citep{LeBorgne2003}, and assuming a Chabrier \citep{Chabrier2003} IMF with lower and upper mass cutoffs of $0.1\,M_{\odot}$ and $100\,M_{\odot}$, respectively.

Using the complete set of SSPs may significantly increase the computational time without necessarily improving the quality of the fit, thus we follow the methodology described in \cite{Dametto2014} to optimise the choice of SSP models that make up the final grid. The final grid contains 210 SSPs spanning 6 metallicities
($Z_{\star}\,[Z_{\odot}] =$ 0.005, 0.02, 0.2, 0.4, 1, 2.5)
and 35 ages
($t\,[\rm{Myr}] =$ 0.13, 0.36, 1.00, 1.66, 2.09, 2.30, 2.40, 2.63, 3.02, 3.16, 3.31, 3.63, 3.98, 5.01, 6.61, 6.92, 7.59, 7.94, 8.32, 9.12, 9.55, 13.8, 20.0, 64.1, 161.0, 255.0, 508.8, 718.7, 1.3$\times$10$^{3}$, 2.0$\times$10$^{3}$, 2.5$\times$10$^{3}$, 4.5$\times$10$^{3}$, 7.0$\times$10$^{3}$, 1.0$\times$10$^{4}$, 1.3$\times$10$^{4}$).

To ensure reliable stellar population synthesis results, the methodology described in this section is applied to model the stellar emission only for clumps with signal-to-noise ratio (S/N) in the featureless continuum (FC) wavelength range of $4730.0\,-\,4780.0$\,\AA\, greater than 10. This (S/N)$_{\mathrm{FC}}$ threshold has been successfully implemented in several previous analyses with \textsc{starlight} \citep[e.g.,][]{Asari2007, Cid2007, Asari2009}. The values of (S/N)$_{\mathrm{FC}}$ for each clump are listed in Table \ref{tab:clump_properties}. In clumps with (S/N)$_{\mathrm{FC}} < 10$ (clumps E, I and J), the continuum is modelled using a median filter with a width of 150\,\AA.  In order to estimate the uncertainties of the derived stellar population properties, we use the spectral variances to draw random samples from a normal distribution and repeat the full-spectral fitting with \textsc{fado} 300 times for each clump spectrum.
Stellar reddening is computed by \textsc{fado}, where we adopt the \citet[][hereafter CCM]{cardelli1989} reddening law. For details on how stellar and nebular attenuation are treated within \textsc{fado}, we refer the reader to Sect.~4 of \citet{Gomes2017}.

\subsection{Emission-line fitting}
\label{subsec:ifscube}
To estimate various gas-phase conditions and physical properties of the clumps, we fit the emission lines observed in their individual integrated spectra. To model the emission-line profiles, we use the \textsc{ifscube}\footnote{\label{foot
}\url{https://ifscube.readthedocs.io/en/latest/}} Python package \citep{ifscube2021}, adopting single Gaussian components and covering the optical wavelength range from 4570 to 7000 \AA. The continuum is modelled using the sum of the best-fit stellar and nebular emission obtained with \textsc{fado}, supplemented by a pseudo-continuum described by a 5th-order polynomial with a 2$\sigma$ rejection threshold to account for residuals from the continuum subtraction. Equivalent widths are computed within an integration radius of five times the velocity dispersion of the line profiles. As constraints, we assume the theoretical line ratios
[\ion{O}{iii}]\,($\lambda$5007/$\lambda$4959)~=~2.98 and
[\ion{N}{ii}]\,($\lambda$6583/$\lambda$6548)~=~3.07 \citep{Storey2000, Osterbrock2006} with the adoption of kinematic groups, i.e. emission lines produced by the same ion species share identical kinematical properties. The latter assumption holds for all emission lines except for H$\beta$ and H$\alpha$, which, due to their spectral separation, may experience different instrumental broadening \citep{Mentz2016,Guerou2017}. To estimate the uncertainties in the fitted parameters, we perform 500 Monte Carlo Markov Chain (MCMC) iterations for each clump. Besides the clump spectra extracted following the methodology described in Sect.~\ref{sec:clump_regions}, we also include in our analysis a spectrum obtained by integration within a circular, PSF-sized aperture centred at the dwarf luminosity-weighted continuum peak, corresponding to the brightest structure of the dwarf seen on the DECaLS colour-composite image shown in Fig.~\ref{fig:MUSE_FoV}. Hereafter, we will refer to this region as the ``nucleus'' of the dwarf.

\begin{figure*}
    \centering
    \includegraphics{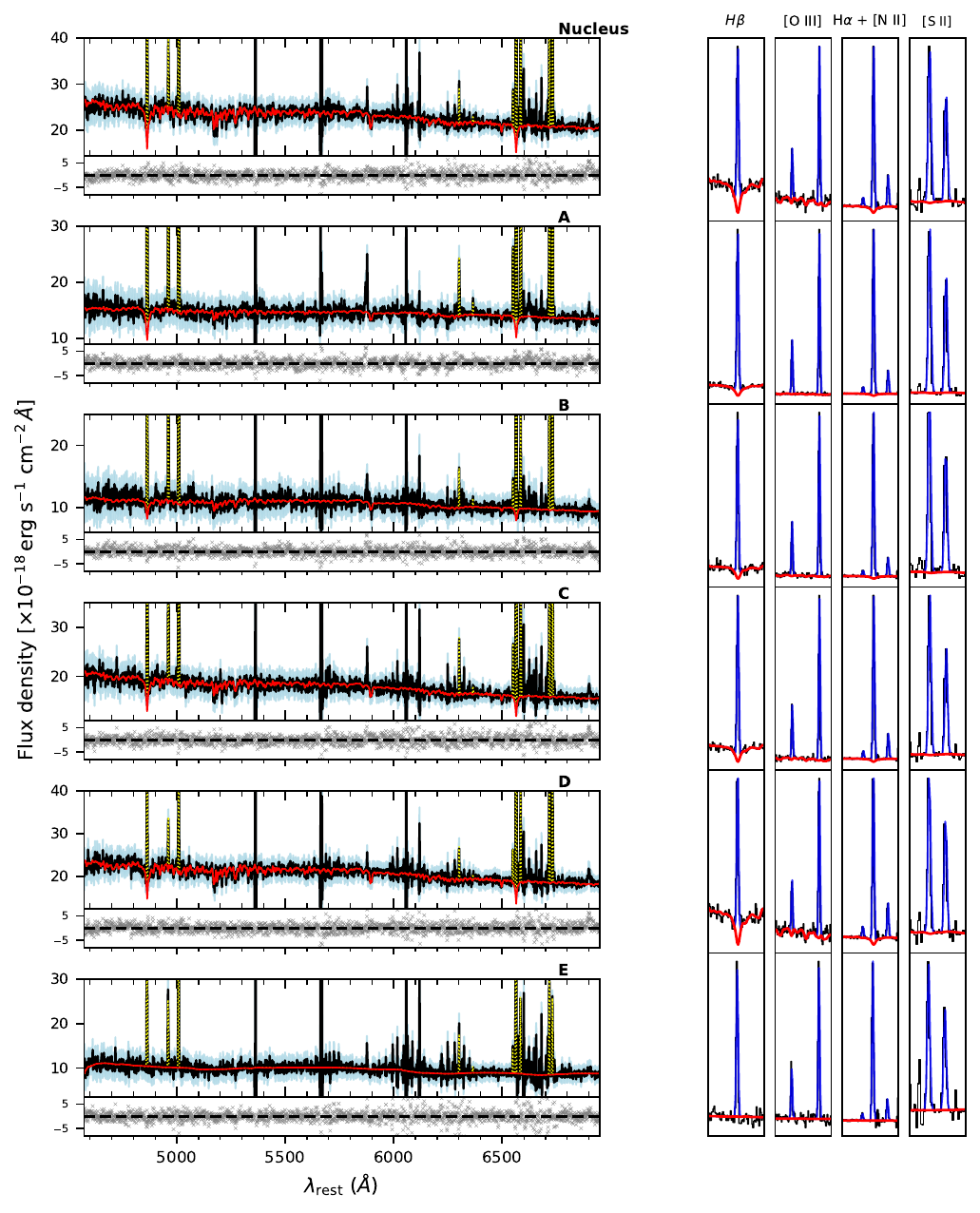}
    \caption{
    \textit{Left panel}: Integrated spectrum (black solid lines) for the ``nucleus'' and clumps A to E. The red line represents the best fit for the continuum that includes the sum of the stellar and nebular emission from \textsc{fado} and the polynomial pseudo-continuum to account for stellar subtraction residuals. The fitted emission lines are presented in yellow. The shadowed region corresponds to the spectral flux density $\pm 3\sigma$. The fit are presented below each spectrum. For spectra with (S/N)$_{\mathrm{FC}} < 10$, the continuum is modelled using a moving median filter with spectral width of 150\,\AA. \textit{Right panel}: Zoom-in of the single Gaussians (blue) fitted to the strongest emission-line profiles of each spectrum.}
    \label{fig:fit_partI}
\end{figure*}

\begin{figure*}
    \centering
    \includegraphics{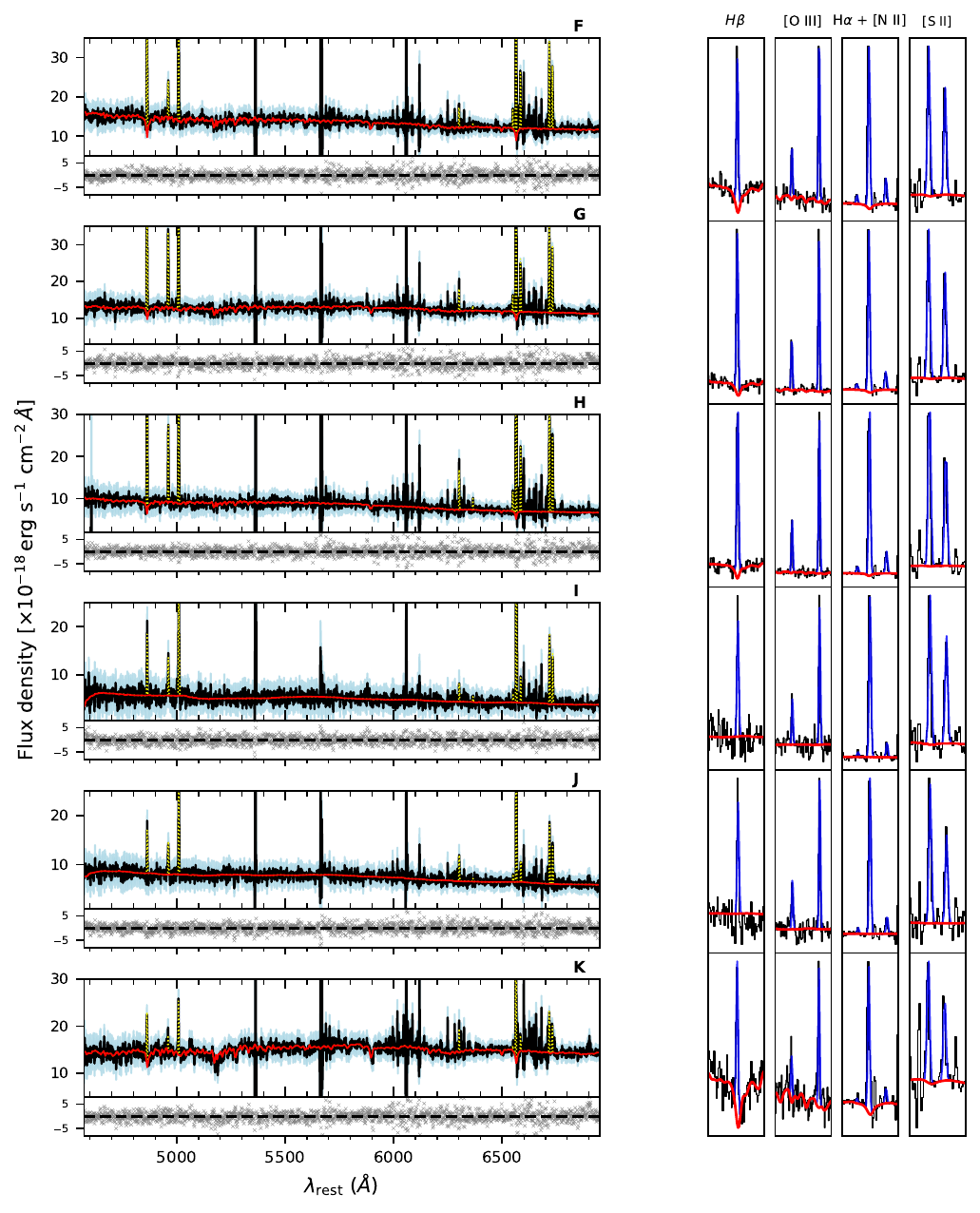}
    \caption{Same as Fig.~\ref{fig:fit_partI}, but for clumps F to J.}
    \label{fig:fit_partII}
\end{figure*}

The spectra and the best fit model are presented in Figs.~\ref{fig:fit_partI} and \ref{fig:fit_partII}. The spectrum is represented by the solid black line, meanwhile the solid red line shows the continuum represented by the addition of the best-fit stellar and nebular emission with the pseudo-continuum. The shadowed region correspond to the $\pm 3 \sigma$ interval. Given that the pseudo-continuum fits only the residuals of continuum subtraction and thus tends to average to $0$ and that the nebular-to-stellar emission is less than 3\% for all clumps, the continuum is approximately the best-fit stellar emission itself. The fitting residuals are shown in grey in the scatter plots at the bottom panels. The estimate of the errors associated with the derived emission-line fluxes follows the prescriptions detailed in \cite{Lenz1992} and \cite{Wesson2016}.


\subsection{Dust attenuation}
\label{subsubsec:dust}
We estimate the dust attenuation of each clump from the Balmer decrement. Assuming the lower limit for electron density, a typical electron temperature of $T_{e} = 10^{4}\,$K for \ion{H}{ii} regions and a CCM reddening law with $R_{\rm{V}} = 3.1$, the expected intrinsic line ratio of H$\alpha$ and H$\beta$ is $(\mathrm{H}\alpha / \mathrm{H}\beta)_{\mathrm{int}} = 2.863$ \citep{Osterbrock2006, PerezMontero2017}. The uncertainty on the colour excess $E(B-V)$ can be estimated by error propagation of the analytical expression:

\begin{equation}
    \delta E(B - V) = \frac{1.086}{k(\lambda_{\mathrm{H}\beta}) - k(\lambda_{\mathrm{H}\alpha})} \ \sqrt{\bigg(\frac{\delta \rm{H}\alpha}   {\rm{H}\alpha}\bigg)^{2} + \bigg(\frac{\delta \rm{H}\beta}{\rm{H}\beta}\bigg)^{2}}
    \label{eq:error_dust}
\end{equation}
 
\noindent where $k(\lambda_{\mathrm{H}\beta})$ and $k(\lambda_{\mathrm{H}\alpha})$ are the values of the CCM reddening curve evaluated at H$\beta$ and H$\alpha$ rest-frame wavelengths, respectively. The terms $\delta \rm{H}\beta$ and $\delta \rm{H}\alpha$ are the errors associated with the flux of H$\beta$ and H$\alpha$ emission line fluxes, respectively.

\subsection{Gas-phase metallicity}
\label{subsubsection:gas_Z}
The most reliable estimates of ionic abundances are obtained using the electron temperature $T_{e}$, derived from the auroral-to-nebular emission line ratios of temperature-sensitive lines, e.g.
[\ion{O}{iii}]~($\lambda \lambda4959,5007$~/~$\lambda 4363$), 
[\ion{N}{ii}]~($\lambda \lambda6548,83$~/~$\lambda 5755$) and
[\ion{S}{iii}]~($\lambda \lambda9069,9532$~/~$\lambda 6312$) \citep{PerezMontero2017}. In many instances, auroral lines are too weak to be detected, so empirical calibrations based on prominent emission lines are frequently used to estimate the gas-phase metallicity. This is also the case of this work, where none of the auroral lines required to apply the direct method can be detected, even when inspecting the integrated clump spectra where the (S/N) is enhanced compared to the spectra at the spaxel resolution. For the sake of comparison with the results obtained in L21 on a spaxel-by-spaxel basis, we calculate the Oxygen abundance of the clumps using the strong-line calibrations from \cite{Marino2013} and \cite{Dopita2016}:

\begin{align}
    12 + \log(\mathrm{O/H})_{\mathrm{M13}} &= 8.533 (\pm 0.012) - 0.214 (\pm 0.012) \times \mathrm{O3N2} \label{eq:marino_metal} \\
    12 + \log(\mathrm{O/H})_{\mathrm{D16}} &= 8.77 + \mathrm{N2S2} + 0.264 \times \mathrm{N2} \label{eq:dopita_metal}
\end{align}

\noindent where the indices O3N2 \citep{Alloin1979}, N2 \citep{Thaisa1994} and N2S2 \citep{Sabbadin1977,Viironen2007} are defined as follows:

\begin{align*}
    \mathrm{O3N2} =& \log \bigg( \frac{[\ion{O}{iii}] \, \lambda 5007}{\mathrm{H}\beta} \times \frac{\mathrm{H}\alpha}{[\ion{N}{ii}] \, \lambda 6583} \bigg) \\
    \mathrm{N2} =& \log \bigg( \frac{[\ion{N}{ii}] \, \lambda 6583}{\mathrm{H}\alpha} \bigg) \\
    \mathrm{N2S2} =& \log \bigg( \frac{[\ion{N}{ii}] \, \lambda 6583}{[\ion{S}{ii}] \, \lambda \lambda 6717,31} \bigg)
\end{align*}

The calibration of Eq.~\ref{eq:marino_metal} is obtained from more than 3000 $T_e$\,--\,based \ion{H}{ii} regions from the Calar Alto Legacy Integral Field Area (CALIFA) survey \citep{Sanchez2012} and is valid through the interval $-1.1 < \mathrm{O3N2} < 1.7$. The calibration of Eq.~\ref{eq:dopita_metal}, on the other hand, is obtained using \textsc{mappings} photoionization code \citep{Sutherland2017, Sutherland2018} to generate a large grid of models from which emission line ratios are used to diagnose the gas-phase metallicity. One of the advantages of this calibration is that it is nearly
independent of variations in the ionisation parameter in the ISM. 

To estimate the uncertainties on the gas-phase metallicity values, we use the errors of the emission-line fluxes obtained from \textsc{ifscube} to draw random samples following a normal distribution and re-compute the metallicities 500 times. Since the emission-line ratios in equations \ref{eq:marino_metal} and \ref{eq:dopita_metal} are close in the spectral domain, we can use the original emission line fluxes to avoid propagating the errors associated with the reddening correction unnecessarily.

\subsection{Star Formation Rates}
The SFR is proportional to the production rate of ionising photons \citep{Kennicutt1998}. Consequently, Hydrogen recombination lines, particularly H$\alpha$, trace the SFR over timescales of $\lesssim$ 10\,Myr, which is the lifetime of the very massive stars responsible for the ionising flux in star-forming (SF) regions, where the SFR is assumed to remain constant within this relatively short timespan. The SFR of the clumps is estimated using the \cite{Kennicutt1998} calibration considering a PADOVA evolutionary track and a Chabrier IMF:

\begin{equation}
    \log\,\mathrm{SFR}\,[M_{\odot}\,\mathrm{yr}^{-1}] = -41.34 + \log L_{\mathrm{H}\alpha, 0}\,[\mathrm{erg}\,\mathrm{s}^{-1}]
    \label{eq:SFR}
\end{equation}

\noindent where $L_{\mathrm{H}\alpha, 0}$ corresponds to the de-reddened H$\alpha$ luminosity of the clumps, determined following the prescriptions detailed in Sect.~\ref{subsubsec:dust}.

\section{Results}
\label{sec:results}

\subsection{Gas rotation pattern}
\label{sec:vel_slices}
\begin{figure*}
    \centering
    \includegraphics[scale=0.9]{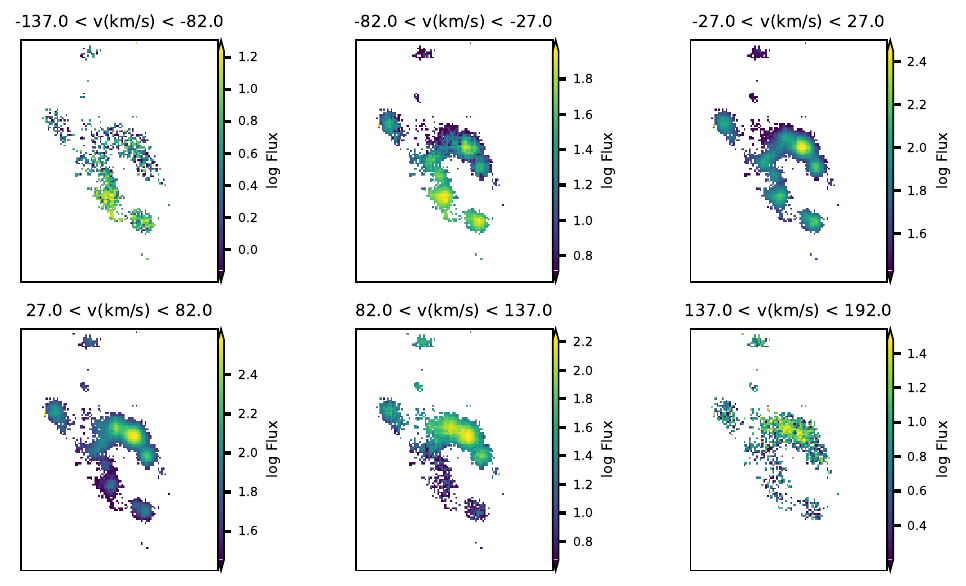}
    \caption{Velocity-sliced maps of the emitting gas in \catname. The velocity intervals covered by each channel are presented on top of each frame. The scale bar represents the logarithm of the flux above the continuum level in units of $10^{-20}$erg\,s$^{-1}$\,cm$^{-2}$\,\AA$^{-1}$. A (S/N)-cut in the spaxels with (S/N)$_{\mathrm{H}\alpha}  < 3$ was applied.}
    \label{fig:velocity_maps}
\end{figure*}

In L21 we presented the map of rotation velocity of the emitting gas, where we observed that the axis of rotation represented by the $v=0$ line crosses the dwarf nucleus shown in Fig.~\ref{fig:MUSE_FoV}. However, the position of the clumps is not clear by only looking at the $v_{\rm{rot}}(\rm{H}\alpha)$ map. Velocity channel maps can be used to further investigate the gas dynamics and to trace the kinematics of the clumps \citep[e.g. see][]{Riffel2006, Genzel2011}. In Fig.~\ref{fig:velocity_maps} we present the behaviour of H$\alpha$ emission-line profile in velocity slices, i.e. given $\lambda_{0}$ as the H$\alpha$ rest-frame wavelength, each frame of Fig.~\ref{fig:velocity_maps} represents a velocity interval with respect to the emission-line centre calculated by the expression $\Delta v = c \Delta \lambda/ \lambda_{0}$. Each channel covers a velocity range of $\sim$~54\,km\,s$^{-1}$, roughly corresponding to the resolution limit of MUSE spectrograph at $\lambda_{0}$. The maps are built considering the continuum-subtracted cube and the colour scale represents the logarithm of the flux above the continuum level, in units of $10^{-20}$erg\,s$^{-1}$\,cm$^{-2}$\,\AA$^{-1}$. A mask was applied to the spaxels with (S/N)$_{\mathrm{H}\alpha}  < 3$.

Noticeably, the flux rapidly drops to the continuum level for velocities exceeding $\sim$~150 km\,s$^{-1}$, but the rotation pattern indicated by the $v_{\rm{rot}}(\rm{H}\alpha)$ map is also observed: The southern part of the dwarf, which includes the clumps G and F is moving towards the observer line-of-sight, whereas the clumps from the upper half of \catname are moving away. Fig.~\ref{fig:velocity_maps} shows that the clumps are located within the rotating component, with the most of them residing preferentially in the northern part of the dwarf. On the other hand, the ``nucleus'' shows little to no evidence of rotation: At $\Delta v \sim$~100 km\,s$^{-1}$, its flux already drops to approximately the continuum-level. The observed rotating structure and the fact that the velocity field of the gas was fairly described by the kinematic model from \cite{Bertola1991}, which assumes gas rotating in circular orbits, reinforce the possibility of a disk component in \catname. If such disk component exists, it would be expected to be subject to the tidal forces arising from the interaction with the companion ETG.

\subsection{Stellar population synthesis}
\label{subsec:SPS_results}

\begin{figure*}
    \centering
    \includegraphics[scale=0.9]{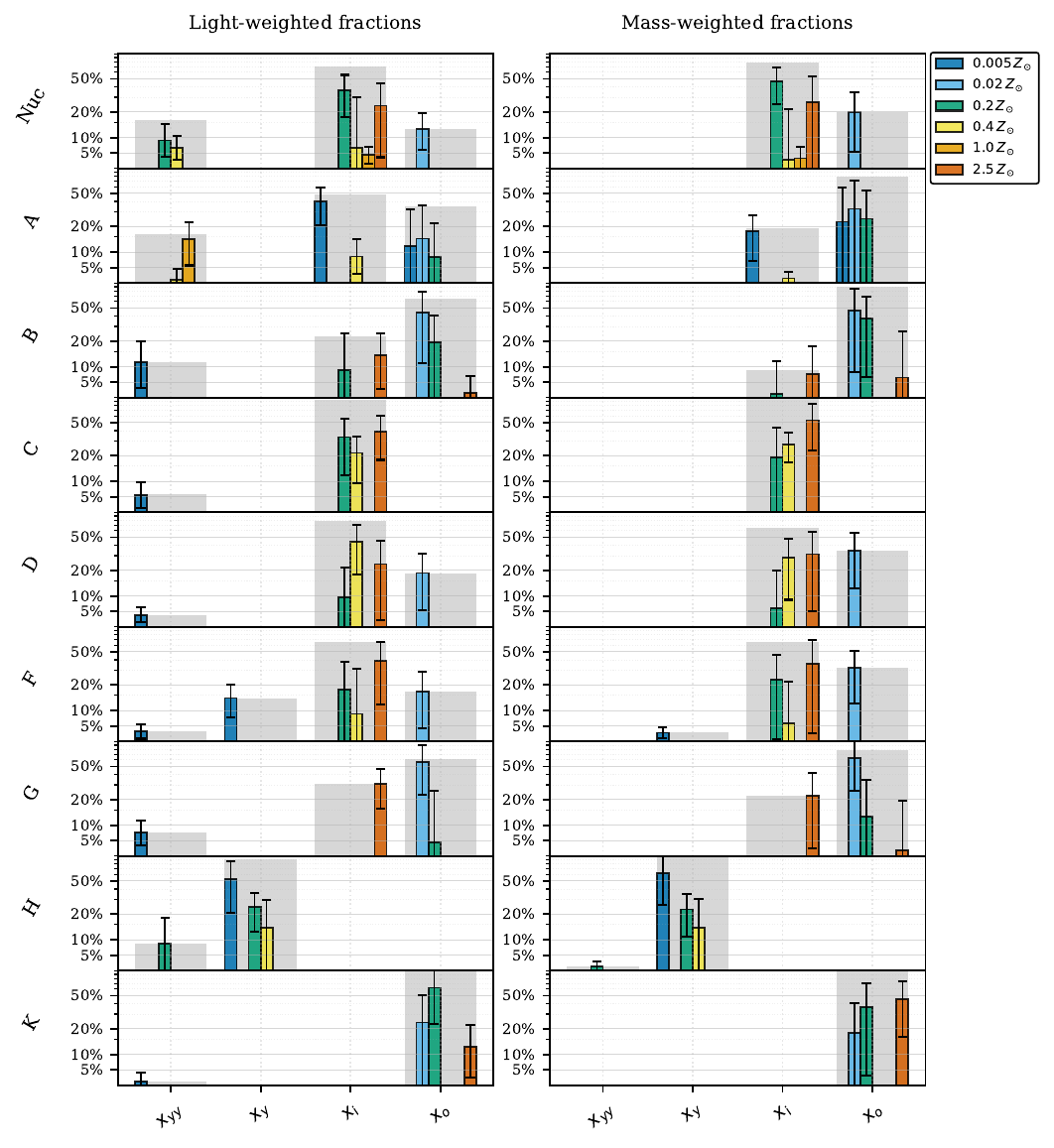}
    \caption{Stellar population synthesis results for the nucleus and SF regions with (S/N)$_{\mathrm{FC}} > 10$. For each region, we present both the luminosity-weighted (right column) and mass-weighted (left column) fractions with respect to the best-fit obtained with \textsc{fado}. The grey bars represent the contributions of each PV grouped into the following age bins: Very young SSPs ($x_{yy} : t \leq 10\,$Myr), Young ($x_{y} : 10\,\mathrm{Myr} < t \leq 100\,$Myr), Intermediate ($x_{i} : 100\,\mathrm{Myr} < t \leq 2\,$Gyr) and Old ($x_{o} : t > 2\,$ Gyr). Thinner bars indicate the contributions within each age bin, coloured according to their stellar metallicities, with the colour scheme presented in the legend at the top right. The $y$-scale is linear up to 20\% and logarithmic for higher fractions, to highlight contributions of underlying sub-populations. Error bars indicate the uncertainty on the contribution of each sub-population, estimated from the 300 MCMC runs.}
    \label{fig:synth_results}
\end{figure*}

In addition to enabling the subtraction of continuum emission and the assessment of gas-phase properties from emission-only spectra, the self-consistent full-spectral fitting with \textsc{fado} can also be used to examine the stellar populations in the SF regions shown in Fig.~\ref{fig:knots}. We present the results of the stellar population synthesis (SPS) for clumps with (S/N)$_{\mathrm{FC}} > 10$ in Fig.~\ref{fig:synth_results}. For each SF region, we display both the luminosity ($x_j$, left column) and mass ($\mu_j$, right column) fractions. We adopt the widely used discrete representation of the Star Formation History (SFH) through SPS \citep[see e.g.,][]{Cid2001, Riffel2011} by grouping the stellar populations in age bins defined as follows:

\begin{itemize}
  \item Very young ($x_{yy}$): $t \leq 10\,$Myr
  \item Young ($x_{y}$): $10\,\mathrm{Myr} < t \leq 100\,$Myr
  \item Intermediate ($x_{i}$): $100\,\mathrm{Myr} < t \leq 2\,$Gyr
  \item Old ($x_{o}$): $t > 2\,$ Gyr
\end{itemize}

The oldest Population Vector (PV) spans a larger range of ages due to the well-known degeneracy between old SSP spectra, while the choice for the youngest PV aims to trace the stellar populations responsible for the ionisation of the gas within the aperture enclosing each clump. While the grey bars in Fig.~\ref{fig:synth_results} correspond to the contributions of each PV weighted by luminosity and mass, the inner bars show the contributions within each PV in terms of $Z_{\star}$, with the colour scheme indicated in the legend at the top right. Error bars show the uncertainties on the contributions of each sub-population, estimated from the 300 MCMC runs. Populations contributing with less than 1\% are neglected in Fig.~\ref{fig:synth_results}.

In the presence of young stellar populations, $x_j$ and $\mu_j$ may differ significantly, as even a small number of OB stars can outshine the other stellar spectral types, thereby dominating the overall light-weighted contributions. On the other hand, these massive stars have shorter lifetimes and are less numerous than their low-mass counterparts, which make up the bulk of the stellar mass and thus dominate the mass-weighted contributions \citep{Hopkins2018}. 
This effect is evident in Fig.~\ref{fig:synth_results}, where $x_j$ is always higher than $\mu_j$ for the $t \leq 100\,$Myr stellar populations.
In fact, transitioning from the light-weighted to the mass-weighted panels shows that decreasing contributions from the younger stellar populations are usually followed by increasing contributions from the older populations, suggesting mass assembly in these regions likely happened at $t >100\,$Myr.
SF region B serves as a clear example of this effect at play. While we measure $x_{yy}$\,[\%] = $12\pm8$ and 
$x_{o}$\,[\%] = $70\pm40$ for the luminosity-weighted case, in the mass-weighted case these values shift to 
$x_{yy}$\,[\%] = $0.06\pm0.05$ and 
$x_{o}$\,[\%] = $91\pm54$, suggesting that stellar mass build-up happened at $t > 2\,$Gyr.

\begin{figure}
    \centering
    \includegraphics[width=\columnwidth]{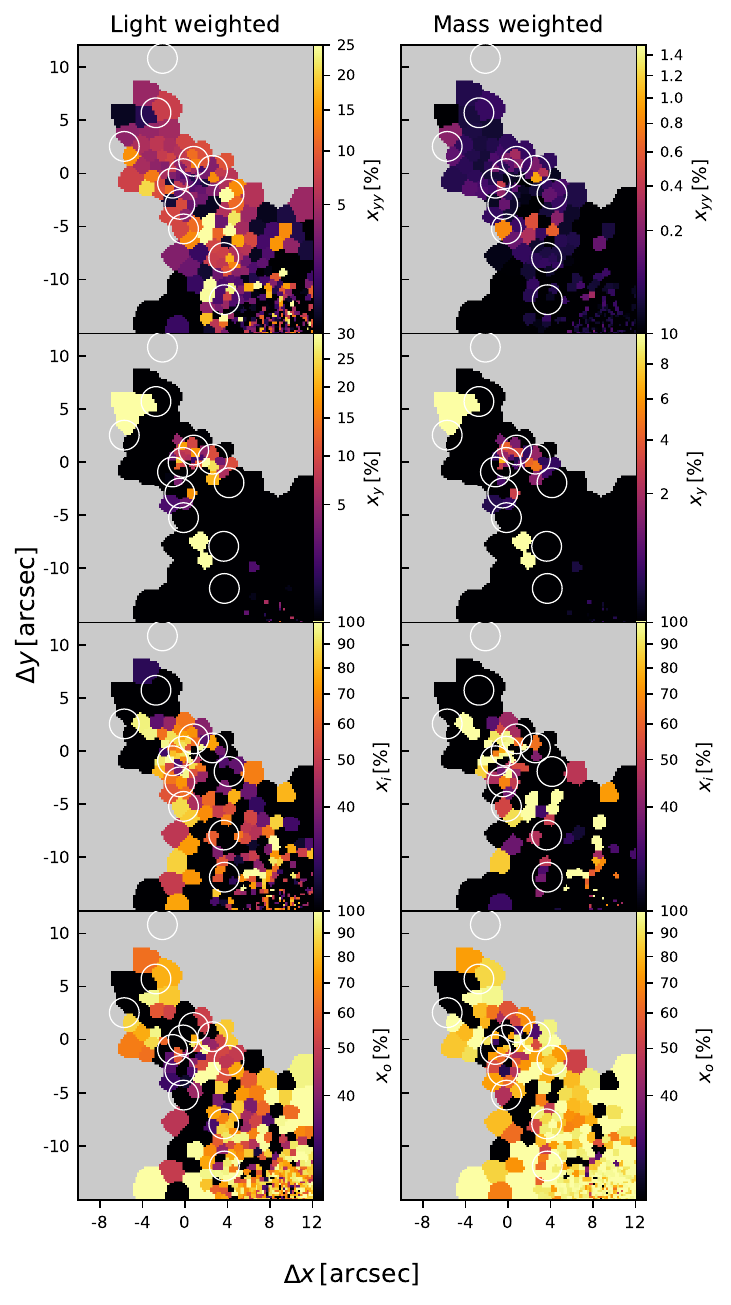}
    \caption{Light- and mass-weighted contributions across \catname, separated into four age bins (definitions on the text). To achieve reliable SPS results, we spatially binned our galaxy with the Voronoi tessellation technique targeting (S/N)$_{\mathrm{FC}}$ = 10. For reference, we also display the PSF-sized circular apertures used for the extraction of the clump spectra detected using \textsc{astrodendro}.}
    \label{fig:galaxy_fracmaps}
\end{figure}

To gain a broader view of how stellar populations are distributed across \catname and how they relate to the identified clumps, we also inspect the spatial distribution of stellar populations within the host galaxy using the same age bins.
To achieve reliable SPS results on the faint continuum of \catname, we spatially bin the galaxy using the weighted adaptation \citep{Diehl2006} of the Voronoi tessellation technique \citep{Cappellari2003} with a target of (S/N)$_{\mathrm{FC}}$~=~10. These maps are presented in Fig.~\ref{fig:galaxy_fracmaps}. The right-hand and left-hand panels show the light- and mass-weighted contributions for each Voronoi bin, respectively. In white, for reference, we overlay the PSF-sized circular apertures used for the extraction of the clump spectra shown in Fig.~\ref{fig:synth_results}.

With the exception of the very first panel, where an increase of the light-weighted contribution from the $t \lesssim 10$\,Myr PV coincides with the position of clump C, the stellar populations appear to be homogeneously distributed, especially the older ones. In other words, the clumpy structure of \catname is not perceptible when examining only the distribution of the $t \lesssim 10\,$Myr stellar populations. Simultaneously, the maps in Fig.~\ref{fig:galaxy_fracmaps} suggest that the dominant stellar populations in \catname have ages above 100\,Myr, with most bins being almost entirely dominated by emission from $t > 2$\,Gyr stellar populations. The implications and interpretations of the results shown in Figs. \ref{fig:synth_results} and Fig.~\ref{fig:galaxy_fracmaps} will be further discussed in Sect.~\ref{sec:discussion}.

\subsection{Clump individual properties}
\label{ref:clump_props}

Several clumpy SF regions are observed in the H$\alpha$ morphology of \catname (see Fig.~\ref{fig:MUSE_FoV}). These regions were separated from the extended ionised gas component using \textsc{astrodendro}, and their spectra were extracted using PSF-sized circular apertures, which are shown in Fig.~\ref{fig:fit_partI} and Fig.~\ref{fig:fit_partII}. Additionally, a spectrum of the continuum peak emission (referred to as the nucleus throughout this paper) was extracted. This section summarises the main properties of these SF regions, determined following the methodology described in Sect.~\ref{sec:sec3} and presented in Table \ref{tab:clump_properties}.

The second column of Table \ref{tab:clump_properties} presents the (S/N)$_{\mathrm{FC}}$ of the clumps, calculated over the 4730\,--\,4780\,\AA\, wavelength range. The signal of \catname is noticeably stronger in the central regions, corresponding to the nucleus and SF regions D and F. Interestingly, SF region C, located farther from the centre than SF regions A and B, exhibits a higher signal. In contrast, SF region E shows a noticeable decrease in signal. However, (S/N)$_{\mathrm{FC}}$ cannot be directly interpreted as a proxy for stellar mass, as demonstrated by the third column, which lists the stellar masses of each SF region derived from the full-spectral fitting with \textsc{fado}. For instance, despite exhibiting higher (S/N)$_{\mathrm{FC}}$, the stellar masses of the nucleus and SF regions A, B, C, and D are nearly identical within the associated uncertainties. Notable deviations from this trend are clumps H, which has a lower stellar mass, and clump K, which shows an intriguing increase in stellar mass towards the galaxy outskirts. We remind the reader that clumps E, I and J have (S/N)$_{\mathrm{FC}} < 10$ and and were therefore excluded from the \textsc{fado} fitting, explaining the absence of SPS-derived properties for these regions in Table \ref{tab:clump_properties}.

\begin{table*}
    \centering
    \caption{Table summarising properties of the star-forming regions detected using \textsc{astrodendro} and the ``nucleus''. The IDs were defined according to a first-order surface F(H$\alpha$) density estimate. The following columns are (i) (S/N) in the featureless continuum 4730\,--\,4780\,\AA. (ii) Stellar mass from \textsc{fado} SPS best fit. (iii) Star formation rate determined using the \protect\cite{Kennicutt1998} relation. (iv) Colour excess determined from the Balmer decrement. (v) Mean age from SPS results. (vi) Mean stellar metallicity from SPS results. (vii) Gas-phase metallicities obtained from strong-line calibrations.}
    \label{tab:clump_properties}
    \begin{threeparttable} 
    \begin{tabular}{cccccccc}
        \toprule
        \toprule
        ID & (S/N)$_{\mathrm{FC}}$ & $\log (M_{\ast} / M_{\odot})$ & $\log\,(\mathrm{SFR}\; [M_{\odot}\,\mathrm{yr}^{-1}])$ & $E(B-V)\;[\mathrm{mag}]$ & $\langle \log t\,[\mathrm{yr}] \rangle$ & $\langle Z_{\ast}\rangle \;[Z_{\odot}]$ & 12 + $\log(\mathrm{O/H})$ \\
        \hline 
        \multirow{2}{*}{Nucleus} & \multirow{2}{*}{23.99} & \multirow{2}{*}{$8.04 \pm 0.07$} & \multirow{2}{*}{$-2.14 \pm 0.01$} & \multirow{2}{*}{$0.02 \pm 0.01$} & $9.1 \pm 0.2$\,\tnote{a} & $0.8 \pm 0.3$\,\tnote{a} & $8.38 \pm 0.01$\,\tnote{c} \\
         &  &  &  &  & $9.2 \pm 0.3$\,\tnote{b} & $0.8 \pm 0.4$\,\tnote{b} & $8.183 \pm 0.008$\,\tnote{d} \\
        \hline 
        \multirow{2}{*}{A} & \multirow{2}{*}{14.76} & \multirow{2}{*}{$8.3 \pm 0.1$} & \multirow{2}{*}{$-1.725 \pm 0.009$} & \multirow{2}{*}{$0.128 \pm 0.009$} & $9.7 \pm 0.2$ & $0.2 \pm 0.1$ & $8.30 \pm 0.02$ \\
         &  &  &  &  & $10.0 \pm 0.1$ & $0.1 \pm 0.1$ & $8.125 \pm 0.005$ \\
        \hline 
        \multirow{2}{*}{B} & \multirow{2}{*}{12.55} & \multirow{2}{*}{$8.0 \pm 0.1$} & \multirow{2}{*}{$-2.04 \pm 0.01$} & \multirow{2}{*}{$0.16 \pm 0.01$} & $9.7 \pm 0.2$ & $0.5 \pm 0.2$ & $8.25 \pm 0.02$ \\
         &  &  &  &  & $9.9 \pm 0.1$ & $0.4 \pm 0.3$ & $8.08 \pm 0.01$ \\
        \hline 
        \multirow{2}{*}{C} & \multirow{2}{*}{20.20} & \multirow{2}{*}{$7.8 \pm 0.1$} & \multirow{2}{*}{$-1.92 \pm 0.01$} & \multirow{2}{*}{$0.072 \pm 0.009$} & $8.6 \pm 0.5$ & $1.1 \pm 0.3$ & $8.34 \pm 0.02$ \\
         &  &  &  &  & $9 \pm 1$ & $1.5 \pm 0.4$ & $8.154 \pm 0.006$ \\
        \hline 
        \multirow{2}{*}{D} & \multirow{2}{*}{25.31} & \multirow{2}{*}{$8.00 \pm 0.06$} & \multirow{2}{*}{$-2.20 \pm 0.02$} & \multirow{2}{*}{$0.03 \pm 0.02$} & $9.2 \pm 0.1$ & $0.8 \pm 0.3$ & $8.39 \pm 0.01$ \\
         &  &  &  &  & $9.4 \pm 0.2$ & $0.9 \pm 0.3$ & $8.21 \pm 0.01$ \\
        \hline 
        \multirow{2}{*}{E} & \multirow{2}{*}{9.25} & \multirow{2}{*}{\ldots} & \multirow{2}{*}{$-2.08 \pm 0.02$} & \multirow{2}{*}{$0.19 \pm 0.02$} & \ldots & \ldots & $8.31 \pm 0.02$ \\
         &  &  &  &  & \ldots & \ldots & $8.13 \pm 0.01$ \\
        \hline 
        \multirow{2}{*}{F} & \multirow{2}{*}{17.54} & \multirow{2}{*}{$7.85 \pm 0.08$} & \multirow{2}{*}{$-2.385 \pm 0.003$} & \multirow{2}{*}{\ldots} & $9.2 \pm 0.2$ & $1.0 \pm 0.4$ & $8.36 \pm 0.02$ \\
         &  &  &  &  & $9.3 \pm 0.3$ & $1.0 \pm 0.4$ & $8.15 \pm 0.01$ \\
        \hline 
        \multirow{2}{*}{G} & \multirow{2}{*}{13.41} & \multirow{2}{*}{$8.0 \pm 0.1$} & \multirow{2}{*}{$-2.20 \pm 0.02$} & \multirow{2}{*}{$0.06 \pm 0.01$} & $9.6 \pm 0.2$ & $0.8 \pm 0.3$ & $8.28 \pm 0.02$ \\
         &  &  &  &  & $9.8 \pm 0.2$ & $0.6 \pm 0.4$ & $8.05 \pm 0.01$ \\
        \hline 
        \multirow{2}{*}{H} & \multirow{2}{*}{11.03} & \multirow{2}{*}{$7.4 \pm 0.2$} & \multirow{2}{*}{$-2.17 \pm 0.02$} & \multirow{2}{*}{$0.09 \pm 0.02$} & $7.8 \pm 0.8$ & $0.1 \pm 0.3$ & $8.31 \pm 0.02$ \\
         &  &  &  &  & $7.8 \pm 0.8$ & $0.1 \pm 0.4$ & $8.05 \pm 0.01$ \\
        \hline 
        \multirow{2}{*}{I} & \multirow{2}{*}{4.67} & \multirow{2}{*}{\ldots} & \multirow{2}{*}{$-2.24 \pm 0.05$} & \multirow{2}{*}{$0.33 \pm 0.05$} & \ldots & \ldots & $8.25 \pm 0.02$ \\
         &  &  &  &  & \ldots & \ldots & $7.83 \pm 0.03$ \\
        \hline 
        \multirow{2}{*}{J} & \multirow{2}{*}{7.85} & \multirow{2}{*}{\ldots} & \multirow{2}{*}{$-2.33 \pm 0.06$} & \multirow{2}{*}{$0.39 \pm 0.06$} & \ldots & \ldots & $8.25 \pm 0.02$ \\
         &  &  &  &  & \ldots & \ldots & $7.89 \pm 0.03$ \\
        \hline 
        \multirow{2}{*}{K} & \multirow{2}{*}{10.92} & \multirow{2}{*}{$8.29 \pm 0.09$} & \multirow{2}{*}{$-2.80 \pm 0.05$} & \multirow{2}{*}{$0.0 \pm 0.05$} & $9.6 \pm 0.2$ & $0.4 \pm 0.3$ & $8.31 \pm 0.02$ \\
         &  &  &  &  & $9.8 \pm 0.2$ & $1.2 \pm 0.8$ & $7.84 \pm 0.05$ \\
        \bottomrule
        \bottomrule
    \end{tabular}
    \begin{tablenotes}
        \item[a] Luminosity-weighted properties.
        \item[b] Mass-weighted properties.
        \item[c] Gas-phase metallicity estimated using strong-line calibration of \cite{Marino2013} (see Eq.~\ref{eq:marino_metal}).
        \item[d] Gas-phase metallicity estimated using strong-line calibration of \cite{Dopita2016} (see Eq.~\ref{eq:dopita_metal}).
    \end{tablenotes}
    \end{threeparttable}
\end{table*}

In the fourth column, we present the SFRs calculated using the \cite{Kennicutt1998} calibration shown in equation \ref{eq:SFR}. Although the regions identified by \textsc{astrodendro} (outlined by the red contours in Fig.~\ref{fig:knots}) span different sizes, the SFR values listed in Table~\ref{tab:clump_properties} can be compared directly, as the spectra were extracted using equally sized circular apertures. SF region A, where the peak of H$\alpha$ emission is located, exhibits the highest SFR, followed by SF regions C and B, corresponding to the northern structure of the rotating gas (see Fig.~\ref{fig:velocity_maps}). Interestingly, the nucleus shows a relatively high SFR despite not appearing in the continuum-subtracted H$\alpha$ map shown in Fig.~\ref{fig:MUSE_FoV}. In the fifth column, we list the colour excess values, estimated from the Balmer decrement. Except for SF region F, all clumps display (H$\alpha$~/~H$\beta$) line ratios greater than or equal to 2.863. Dust reddening appears to be $\sim0.1\,$mag higher in SF regions A and B compared to the central regions, where it is nearly absent. The higher dust reddening values are observed in the northernmost faint clumps, I and J.

In the sixth and seventh columns we list the mean stellar ages and metallicities obtained from \textsc{fado} full-spectral fitting, for both luminosity and mass-weighted cases, defined as follows:

\begin{align*}
<\log t_{\star}>_{L} = \sum\limits_{j}^{N_{\star}} x_{j}\,\log(t_{j}) &\qquad <\log t_{\star}>_{M}= \sum\limits_{j}^{N_{\star}} \mu_{j}\,\log(t_{j})\\
<Z_{\star}>_{L} = \sum\limits_{j}^{N_{\star}} x_{j}\,Z_{j} &\qquad <Z_{\star}>_{M}= \sum\limits_{j}^{N_{\star}} \mu_{j}\,Z_{j}
\end{align*}

\noindent where $N_{\star}$ is the number of SSPs used in the SPS. $t_j$ and $Z_j$ are the age and metallicity of the $j$-th SSP in the grid, respectively. In terms of age, both cases agree fairly well for all clumps, with the mass-weighted mean ages at most 0.4\,dex higher (as expected) than the luminosity-weighted ones.

From Table \ref{tab:clump_properties}, we see that several clumps, such as C, D, and F, as well as the nucleus of \catname, show mean ages of approximately 1\,Gyr. Interestingly, clumps A and B exhibit older mean ages of $\gtrsim 6\,$Gyr.
As shown in Fig.~\ref{fig:synth_results}, most clump spectra exhibit higher fractions of intermediate and old stellar populations, which consequently dominate the average age values. Therefore, the reported mean age values do not trace clump ages, but rather reflect the underlying older stellar populations, likely associated with the disk of \catname.
Notably, clump H stands out with a lower mean age of approximately 60\,Myr, despite the highest associated age uncertainties. This result is consistent with Fig.~\ref{fig:synth_results}, where clump H is the only SF region lacking any contribution from $t \gtrsim 100\,$Myr stellar populations, regardless of whether light- or mass-weighted quantities are considered.

In terms of stellar metallicities, the results reported in Table \ref{tab:clump_properties} reflect highly mixed older populations seen in Fig.~\ref{fig:synth_results}.
Clumps A and B show mean stellar metallicities ranging from $0.1\,Z_{\odot}$\,--\,$0.5\,Z_{\odot}$, whereas other clumps located near the central region of \catname, such as clumps D and F, consistently suggest solar metallicities.
Clump K stands out as a particularly striking example. Its mean stellar metallicity shifts from $0.4\,Z_{\odot}$ in the light-weighted case to $1.2\,Z_{\odot}$ in the mass-weighted case. As seen in Fig.~\ref{tab:clump_properties}, this occurs because the contribution from a $Z_{\star} = 2.5\,Z_{\odot}$ subpopulation increases from a light fraction of roughly 15\% to a mass fraction of nearly 50\%. However, the associated uncertainties are substantial, which is expected given that clump K is located at the detection limit of the continuum emission in \catname.

Overall, the values of Z$_{\mathrm{gas}}$ in the clumps agree well with the galaxy mean value of 12 + log(O/H) = 8.2 reported in L21, when considering the strong-line calibration from \cite{Marino2013}. However, they are, on average, approximately 0.2\,dex when the calibration from \cite{Dopita2016} is considered. For the more distant clumps I, J and K, this discrepancy becomes more pronounced, with clump K reaching a difference of nearly 0.5\,dex.

\section{Discussion}
\label{sec:discussion}

In this section, we investigate the potential contamination of the integrated continuum emission of clumps G and K by the extended luminous halo of Mrk~1172. We also explore the scenarios of in situ and ex situ clump formation in \catname, in light of the results presented in Sect.~\ref{sec:results}, and discuss the physical implications of the observed flat distribution of gas-phase metallicities across the galaxy.

\subsection{Contamination from Mrk~1172 luminous halo on clump spectra}
\label{subsec:contamination}

A visual inspection of Fig.\ref{fig:MUSE_FoV} suggests a possible overlap between the southernmost regions of \catname and the faint northern outskirts of Mrk~1172 luminous halo.
For example, using the isophote levels as a reference, it is possible to observe that the southernmost clumps, G and K, seem to lie in a transition zone between the two galaxies.
Additionally, considering that clump E is located closer to \catname central regions than clump K, it is counter-intuitive that the latter satisfies the minimum (S/N)$_{\mathrm{FC}}$ criterion while the former does not.
Although mild, the intriguing increase of (S/N)$_{\mathrm{FC}}$ in the external parts of the disk is asymmetrical, as both clumps J and I, located at the northernmost side of \catname, exhibit the lowest (S/N)$_{\mathrm{FC}}$ values.
Another particularity of clump K can be seen in Fig.~\ref{fig:fit_partII}. In the zoom-in panel showing the H$\beta$ emission-line profile, the underlying absorption feature associated with the stellar component exhibits a mild but noticeable velocity offset relative to the ionized gas.
In addition to the shifted absorption feature, clump K also exhibits one of the highest stellar mass among the clumps listed in Table~\ref{tab:clump_properties}. According to the SPS results shown in Fig.~\ref{fig:synth_results}, this stellar mas was formed more than 2\,Gyr ago.

A possible explanation for the peculiarities listed above is that the PSF-sized circular aperture used to extract the spectrum of clump K captures part of the light from the outer luminous halo Mrk~1172. If we assume that the stellar emission in clump K is intrinsically faint, any flux contribution from a background or foreground source could dominate the continuum emission, thereby explaining why (S/N)$_{\mathrm{FC}}$ rises slightly in clump K. 
Given the slight redshift difference between the two galaxies, the contamination hypothesis could also account for the velocity offsets between stellar and gas components observed in the spectrum of clump K. Indeed, recent work by \cite{Jegatheesan2025} reports contamination from the dwarf in the outer halo of the ETG.
In this subsection, we investigate whether such contamination is sufficient to dominate the continuum emission in clump K -- or even in other clumps -- and, consequently, be responsible for both the shifted absorption features observed in Fig.~\ref{fig:fit_partII} and the presence of old stellar populations shown in Fig.~\ref{fig:synth_results}.

SPS can be used to test the possibility of contamination. If the continuum emission from these clumps is contaminated by light from the ETG outer halo, their inferred stellar population properties should resemble those of Mrk~1172 rather than those intrinsic to \catname. To test this, we analyse two representative spectra: one extracted from the ETG outer halo, and another from a region of \catname without apparent overlap with Mrk~1172. The procedure is described below.

\begin{figure}
    \centering
    \includegraphics[width=1.04\columnwidth]{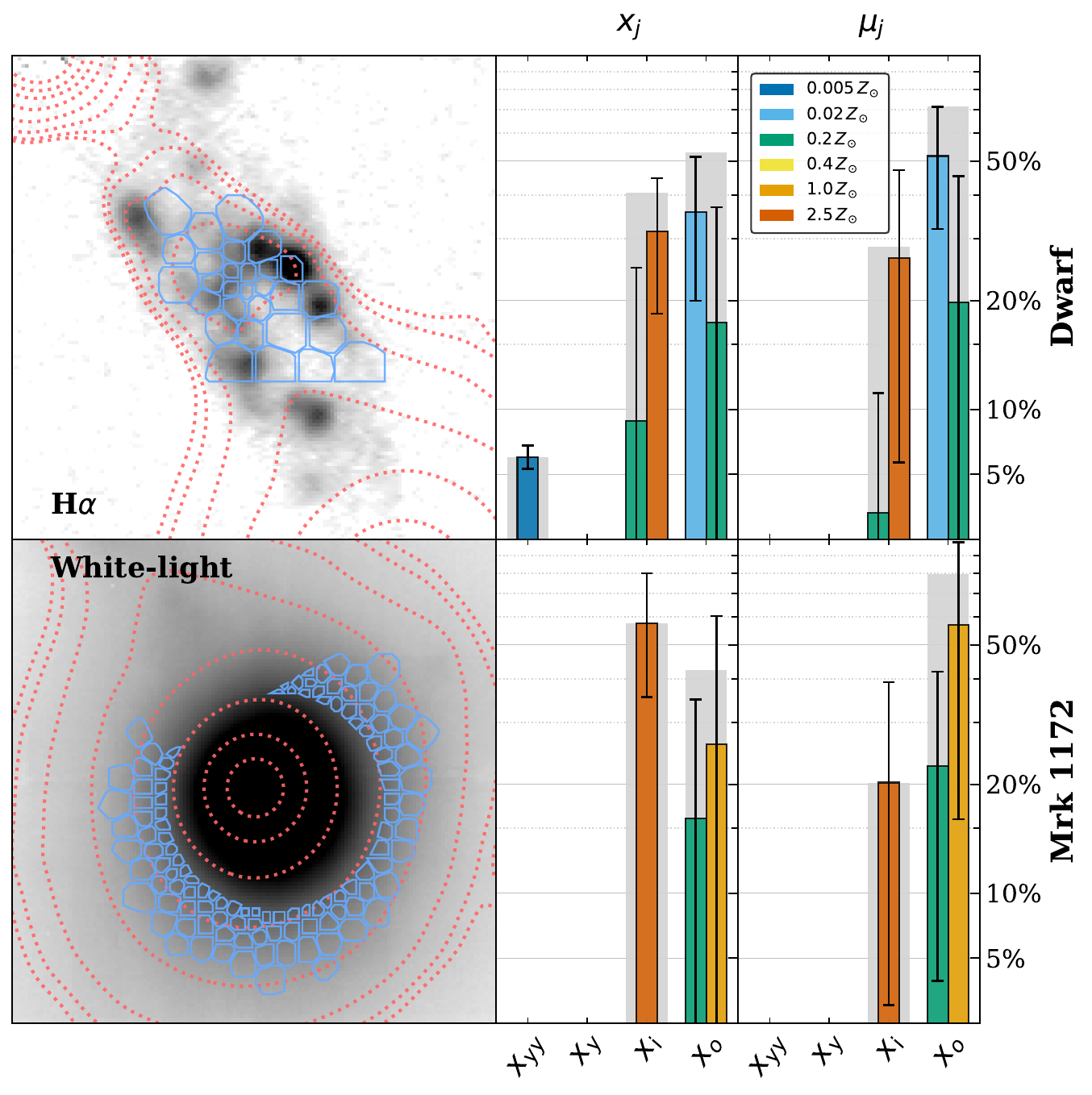}
    \vspace{0.4cm}
    \raggedright
    \includegraphics[width=\columnwidth]{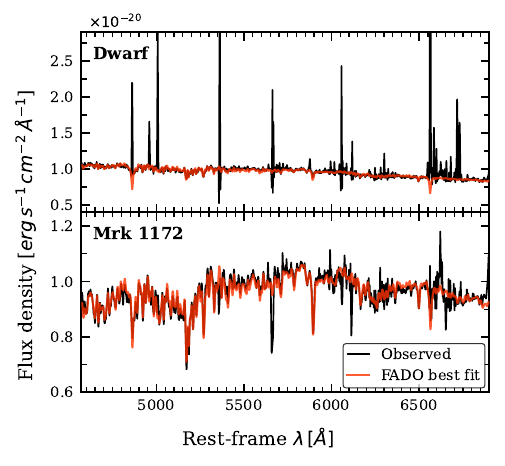}
    \vspace{-1cm}
    \caption{\textit{Top}: Zoom-in images of the dwarf galaxy (H$\alpha$) and Mrk~1172 (white-light image). Red dotted contours indicate isophote levels, in the range $17.9 \lesssim \mu_{\mathrm{AB}, z}\,[\mathrm{mag/arcsec}^{2}] \lesssim 21.5$. The blue polygons show the Voronoi binning patterns, targeting (S/N)$_{\mathrm{FC}}$~=~15, used to extract median stacked spectra for both \catname and the outer luminous halo of Mrk~1172. Regions in which both galaxies overlap were masked. The left histograms display the light-($x_j$) and mass-weighted ($\mu_j$) contributions of both galaxy regions, divided into four age bins: Very young SSPs ($x_{yy} : t \leq 10\,$Myr), Young ($x_{y} : 10\,\mathrm{Myr} < t \leq 100\,$Myr), Intermediate ($x_{i} : 100\,\mathrm{Myr} < t \leq 2\,$Gyr) and Old ($x_{o} : t > 2\,$ Gyr). \textit{Bottom}: Median observed stacked spectra (black) in both regions and their best-fit stellar emission (red) derived from \textsc{fado}, which is used to derive the SFHs.}
    \label{fig:contamination}
\end{figure}

First, we aim to characterise the underlying intermediate and old stellar populations in \catname, possibly related to the disk. Second, we want to characterise the stellar populations in the outer halo of Mrk~1172, excluding the region where the two galaxies overlap. 
To achieve the former, we apply a Voronoi binning to \catname targeting (S/N)$_{\mathrm{FC}}$~=~15, keeping clumps G and K masked.
For the latter, we mask both \catname and the overlapping region, and then apply Voronoi binning with the same (S/N)$_{\mathrm{FC}}$~=~15 requirement.
After binning, we also mask the central region of Mrk~1172, as our focus is solely on its outer halo \citep[For a spatially resolved description of the stellar populations of Mrk~1172, we refer the reader to][]{Jegatheesan2025}. The resulting binning patterns are shown in Fig.~\ref{fig:contamination} by the blue regions, overlaid on both the white-light image (bottom panel) and the continuum-subtracted H$\alpha$ image (top panels).
To obtain a representative spectrum for each region, we stack all spectra within the blue areas shown in Fig.~\ref{fig:contamination}. The resulting stacked spectra, along with their best fits provided by \textsc{fado}, are shown in the bottom panel of Fig.~\ref{fig:contamination}. The corresponding light- and mass-weighted contributions for both regions are shown in the left panel of the same figure, using the same colour-coding scheme as in Fig.~\ref{fig:synth_results}.

When integrating the light across \catname without optimizing for clump detection, we find that mean contribution from $t \lesssim 100\,$Myr stars is only $\sim$5\% in the luminosity-weighted case. Since the dwarf is relatively faint in the continuum, the lack of prominent stellar absorption features could lead the SPS code to favor older stellar populations, as these would likely fit better a nearly flat continuum. However, as shown in Fig.~\ref{fig:contamination}, at least one stellar feature is clearly visible in the stacked spectrum of \catname: the wings around the H$\beta$ emission line.

Moreover, the light- and mass-weighted fractions of the intermediate and old populations of \catname shown in Fig.~\ref{fig:contamination} are very similar to those found in in central clumps such as B, C, D, and F, as well as in the nucleus of \catname.
Many of these clumps also exhibit significant fractions of a $Z_{\ast} = 0.4\,Z_{\odot}$ in their intermediate-age stellar populations, a subpopulation that is not detected in the median stacked spectra shown in Fig.~\ref{fig:contamination}.
Interestingly, a very metal-rich population appears to have formed in \catname during the age interval $100\,\mathrm{Myr} \lesssim t \lesssim 2\,$Gyr. The corresponding metal-rich SSPs identified by the SPS suggest this event occurred roughly 1\,Gyr ago.
The recent SF episode in \catname seems to have formed metal-poor stars, but the amount of mass formed is negligible.
The oldest age bin in most of the listed clumps also resembles the stacked case, consistently indicating the presence of $Z_{\ast} = 0.02\,Z_{\odot}$ populations.
Clump K stands out as a notable exception, showing a significant contribution from very metal-rich stars in its oldest population, particularly in the mass-weighted case. This detection is likely due to contamination from the outer halo of Mrk~1172. Although the stellar metallicities do not match exactly, the highly degenerate solutions make direct comparisons between subpopulations highly uncertain, as shown in  Appendix~\ref{appendix:SPS_caveats}.

Given the consistency with which the stellar populations described above are found throughout \catname, it is reasonable to assume that they represent the underlying stellar component of the galaxy, upon which recent star formation has occurred. Notably, the intermediate-age stellar populations in \catname are similar to those found in the outer layers of Mrk~1172. This resemblance may suggest that material has been exchanged between the two galaxies, possibly as a result of the dwarf being continuously disrupted by tidal forces while orbiting its massive companion. Determining whether the timescales required for this speculative scenario are compatible with the $\sim$1\,Gyr event suggested by the SPS analysis would require a dedicated investigation using numerical simulations, which is beyond the scope of this work.

\subsection{Metal-poor gas accretion as the trigger of clumpy SF}
\label{subsec:MP_accretion}
Continuous accretion of metal-poor cold gas was proposed to explain the inverted metallicity gradients (i.e. metallicity decreasing towards the centre) observed in $z\sim$~3 galaxies \citep{Cresci2010}. Simulations indicate that the tidal torques caused by interactions are able to channel significant amounts of gas from the outskirts to the central regions of the galaxy \citep{Muratov2015}. Accretion of cold gas is hard to detect directly due to the faint absorption
features and low densities of the gas \citep{Sanchez-Almeida2014b}, but indirect evidence
is provided by off-center star-forming regions with increased Star Formation Rates and
associated localized drops in the gas-phase metallicity, observed for star-forming dwarf
galaxies \citep{Sanchez-Almeida2014a, Macarena2023}, tadpole galaxies in the local Universe \citep{Sanchez-Almeida2013} and at $z \gtrsim 2$ \citep{Elmegreen2007, Elmegreen2010}, several spiral galaxies within the MaNGA survey \citep{Sanchez-Menguiano2019, Hwang2019} and in extremely metal-poor star-forming galaxies \citep{Papaderos2008}, being also predicted by Adaptive Mesh Refinement (AMR) cosmological simulations \citep{Ceverino2016}. For a detailed review on the processes
related to the accretion of cold gas, we refer to
\cite{Tumlinson2017}. In cases where metal-poor gas fuels star formation, an anti-correlation between gas-phase metallicity and SFR is expected to be observed \citep{Sanchez-Almeida2018, Macarena2023}. 

In L21 we presented single-spaxel resolution spatially resolved maps of the O abundance in \catname using the strong-line calibrations presented in Sect.~\ref{subsubsection:gas_Z}, but instead of observing metallicity drops associated to the position of the clumps, a fairly flat metallicity distribution was observed. However, the metal-poor young stellar populations in the clumps indicated by the SPS results shown in Fig.~\ref{fig:synth_results} motivate us to inspect the relation between gas-phase and $\log\,\Sigma_{\mathrm{H}\alpha}$. To ensure a minimum (S/N) level, we bin our data using the weighted Voronoi tessellation targeting a (S/N)~=~5 in [\ion{N}{ii}]~$\lambda$6583, the weakest emission line necessary to calculate the O abundance using equations \ref{eq:marino_metal} and \ref{eq:dopita_metal}. We allow a maximum cell area of 3\,arcsec$^{2}$, resulting in 286 bins. The continuum of each bin was fitted with \textsc{fado} and the emission lines were modelled with \textsc{ifscube} according to the prescriptions detailed in Sect.~\ref{subsec:ifscube}.

\begin{figure}
    \centering
    \includegraphics[width=\columnwidth]{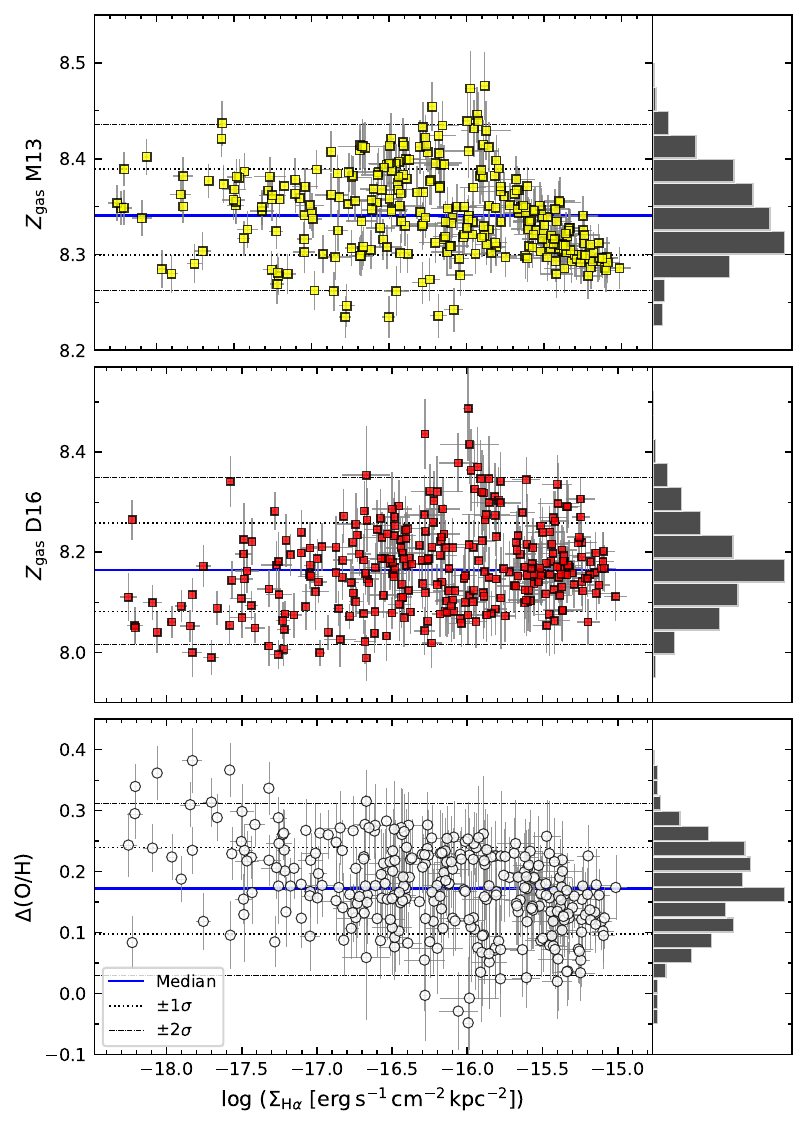}
    \caption{Gas-phase metallicity values as a function of the surface density H$\alpha$ flux ($\Sigma_{\mathrm{H}\alpha}$). Each point represents measurements from Voronoi bin spectra, derived from a tessellation pattern targeting (S/N) = 5 in [\ion{N}{ii}]~$\lambda\,$6583. In the upper panel, $Z_{\mathrm{gas}}$ values were obtained using \,\protect\cite{Marino2013} calibration, while in the middle panel the $Z_{\mathrm{gas}}$ values were derived using the calibration of \,\protect\cite{Dopita2016}. The bottom pannel shows the residuals between both calibrations for each bin. The right-hand panels show the distributions of metallicity values and residuals. The solid blue lines represent the medians of the point distributions, while the dotted and dash-dotted lines correspond to the 15.9\%, 84.1\% and 2.3\%, 97.7\% percentiles of the distribution, respectively.}
    \label{fig:metalpoor_gas_accretion}
\end{figure}

In Fig.~\ref{fig:metalpoor_gas_accretion} we present the gas-phase metallicity values for the extracted bin spectra. The yellow squares in the upper panel show the distribution of $Z_{\mathrm{gas}}$ values obtained using the calibration from equation \ref{eq:marino_metal}, while the red squares represent the same spectra, but with $Z_{\mathrm{gas}}$ values derived from equation \ref{eq:dopita_metal}. In the bottom panel, we show the difference between the metallicity values derived using both calibrations. The values of $\Sigma_{\mathrm{H}\alpha}$ are obtained considering the physical area of each Voronoi bin. Using the cosmological parameters detailed in Sect.~\ref{sec:intro} and the estimated redshift of \catname, each spaxel corresponds to a physical area of approximately 0.03\,kpc$^{2}$.

For both calibrations, fairly flat metallicity distributions are observed, consistent with the metallicity maps shown in L21. In the upper panel, the metallicity values within $\pm 2\sigma$ of the distribution span approximately 0.2\,dex, whereas in the middle panel it is approximately 0.35\,dex. In the upper panel, a slight negative inclination in the relation becomes noticeable for $\log \Sigma_{\mathrm{H}\alpha} \gtrsim -15.7$, likely corresponding to the locations of the SF clumps. It is worth noting, however, that this potential anti-correlation for $\log \Sigma_{\mathrm{H}\alpha} \gtrsim -15.7$ spans a narrow metallicity range of $\sim$0.1\,dex, and is not present in the middle panel. Despite the absolute $Z_{\mathrm{gas}}$ values derived from both calibrations being systematically offset by $\sim$0.2\,dex on average, the flat distribution is evident in both cases. The average $Z_{\mathrm{gas}}$ values are lower compared to the mean stellar metallicities seen in some clumps such as C, D and F, for instance. On the other hand, they are considerably higher compared to the stellar metallicities of clumps like A and H. 

We stress that the observed flat $Z_{\mathrm{gas}}$ distribution alone is not sufficient to rule out the hypothesis that clump formation is driven by the accretion of metal-poor gas.
As previously mentioned, at the distance of \catname, individual MUSE spaxels correspond to a physical area of 0.03\,kpc$^{2}$, equivalent to a square region of approximately 165\,pc\,$\times$\,165\,pc. In case of any metallicity inhomogeneity, this physical scale is sufficiently large for the dillution of metallicity inhomogeneities to become non-negligible, and consequently potential metallicity drops considerably smaller than the kpc-sized PSF of our observations could be smeared out to the average metallicity of the underlying extended ionised gas component. Such averaging effect tends to become more pronounced with increasing aperture size used to probe the galaxy ISM, as it is the case of the Voronoi bins and the circular PSF-sized apertures used to obtain the gas-phase and stellar metallicities, respectively. This likely explains why the light-weighted SSP contributions in all clumps are dominated by stellar populations older than 10 Myr, despite their selection being based on the H$\alpha$ morphology of \catname. It might also account for the slight anti-correlation observed for the highest H$\alpha$-emitting bins in Fig.~\ref{fig:metalpoor_gas_accretion}, although it does not explain why such behaviour is observed only when the \cite{Marino2013} diagnostic is employed.

Most of the clumps have sizes comparable to the MUSE kpc-sized PSF, suggesting that these size estimates are mostly influenced by seeing limitations. For example, in Appendix~\ref{appendix:GSAOI} we show high-resolution imaging of \catname in the J and \Ks NIR bands, obtained from GSAOI observations. The high spatial resolution of GSAOI provides images with enhanced seeing (approximately 7$\times$ better than MUSE) that indicate the sizes of the clumps could be significantly smaller than the ones inferred from IFU observation. Unfortunately, due to the faint continuum emission of \catname, only clump A and the nucleus were detected with counts at least 3$\times$ above the background level (see Fig.~\ref{fig:GSAOI}). Although the high uncertainties in the faint-end of the luminosity profile of SF region A prevent accurate estimates of its effective radius, Fig.~\ref{fig:GSAOI} suggests a radius of approximately 200\,pc, which is about 5$\times$ smaller than the MUSE PSF.

\subsection{Flat metallicity distributions and in situ clump formation}
\label{sec:flat_metals}

Gas-phase metallicity distributions are commonly calculated in terms of radial profiles to infer metallicity gradients. Although in Fig.~\ref{fig:metalpoor_gas_accretion} the metallicity distribution is presented as a function of $\Sigma_{\mathrm{H}\alpha}$, re-arranging the x-axis to radial distances would likely have little effect on the tightly clustered metallicity values around the mean. Furthermore, as previously mentioned, spatially resolved maps of gas-phase metallicity were investigated in L21, with no evidence found for significant metallicity gradients in \catname. Given that the flat distribution in Fig.~\ref{fig:metalpoor_gas_accretion} is observed across a relatively large range of $\Sigma_{\mathrm{H}\alpha}$, this characteristic corresponds to the extended ionized gas in \catname, on top of which the clumpy SF seems to occur.

Flat metallicity gradients are common among low-mass irregular galaxies \citep{Kobulnicky1997, vanZee2006, Matteucci2012, Bresolin2019}. Recent low-resolution JWST/NIRSpec analysis of three $6 \lesssim z \lesssim 8$ galaxies with masses $7.6 \lesssim \log(M/M_{\odot}) \lesssim 9.3$ also found evidence of flat metallicity gradients \citep{Venturi2024}. Several mechanisms can produce flat gradients, including violent inflows of gas caused in pair collisions and mergers \citep{Kewley2010, Torres-Flores2014}, as well as large-scale processes capable of efficiently mixing and redistributing the gas within the disk \citep{Bresolin2012}. For example, enriched material expelled by strong Supernovae-driven winds and outflows can be re-accreted later \citep{Tenorio-Tagle1996, Gibson2013, Venturi2024}. Simulations also indicate that turbulence driven by gravitational instabilities can facilitate metal mixing within the galaxy \citep{Petit2015}.

The ionized gas velocity map shown in L21 suggests the presence of a disk in \catname, and in Sect.~\ref{sec:vel_slices} we demonstrated that the clumps are part of this rotating gas component. One scenario for the formation of massive clumps involves galactic-scale gravitational instabilities caused by gas inflows onto an existing disk \citep{Agertz2009, Dekel2009a, Bournaud2009}.
However, if the kpc-sized clumps formed in situ, the $t < 10$\,Myr stellar populations responsible for the high H$\alpha$ fluxes in these regions would be expected to form from the surrounding gas and therefore exhibit metallicities similar to those of the host galaxy.

Gas-phase metallicity shows a remarkably flat distribution, which could support the scenario described above. However, a mild ($\sim$0.1\,dex), yet systematic, decrease in metallicity is observed in Fig.~\ref{fig:metalpoor_gas_accretion} when the M13 strong-line calibration is applied. Simultaneously, SPS results suggest that, except for clump A, recent star formation in the kpc-sized clumps of \catname has produced stars with metallicities lower than both the intermediate-age stellar populations and the average $Z_{\mathrm{gas}}$. This supports the hypothesis that their formation was triggered by the accretion of pristine gas.
Notably, this interpretation does not require a large amount of infalling material. As indicated in the mass-weighted panels of Fig.~\ref{fig:synth_results}, the stellar mass formed in these clumps over the past $t \lesssim 10\,$Myr is negligible. We emphasize, however, that determining these metallicities precisely is challenging due to the highly degenerate nature of the SPS solutions, as shown in Appendix~\ref{appendix:SPS_caveats}. Otherwise, it would be difficult to conceive how gas with $0.005\,Z_{\odot}$ could have formed stars without mixing with the ISM of \catname.
SPS might also offer an insight on the observed flat gas-phase metallicity profile of \catname. As discussed in Sect.~\ref{subsec:contamination}, the dwarf appears to have undergone a SF event around 1\,Gyr that formed metal-rich stars, potentially contribution to the overall chemical homogenization of its ISM.

The interacting system composed by the ETG Mrk~1172 and \catname is fascinating, with still many open questions. High-resolution imaging probing both the rest-frame NUV and NIR emission can help disentangling the properties of the clumps from the underlying disk. At the same time, hydrodynamical simulations informed by the observational constraints reported in this work and in L21 can be extremely valuable to shed led into the role of the ETG on the evolution of \catname, as well as the timescales involved in these processes.

\section{Summary and conclusions}
\label{sec:conclusion}
This work aims to extend the analysis of the dwarf irregular galaxy \catname first presented in \cite{Lassen2021}. The dwarf is in interaction with the massive \citep[$\rm{log} (M/M_{\odot}) \approx 11.0$,][]{Omand2014,Lassen2021} Early-Type Galaxy Mrk~1172. While L21 focused on general properties of the dwarf through single-spaxel spatially resolved analysis, the present work aims to constrain the physical and chemical properties of a few kpc-sized star-forming clumps that are prominent in \catname when observed in H$\alpha$. We investigate mechanisms that might have triggered clumpy star formation in this galaxy, including the infall of pristine gas and clump formation within gravitationally unstable regions of the disk. Below we summarise our conclusions.

\begin{itemize}
    \item The clumpy morphology of \catname becomes evident only when observed in H$\alpha$. In the optical continuum, the morphology of the dwarf is dominated by two compact structures surrounded by faint tidal features. These sources correspond to the luminosity-weighted continuum peak (referred to as the nucleus throughout this work) and star-forming (SF) region A, which represents the peak of star formation rate (SFR) in \catname. When viewed in continuum-subtracted H$\alpha$, the nucleus is not detected.

    \item \textsc{astrodendro} was applied to identify clumps and separate them from the underlying extended ionized gas component of \catname. Clump spectra were extracted using PSF-sized circular apertures, allowing us to derive several properties. They present SFRs in the range $-1.7 \lesssim \log (\mathrm{SFR}\,[M_{\odot}\,\mathrm{yr}^{-1}]) \lesssim -2.8$. For clumps with sufficient signal, we employed \textsc{fado} to perform a self-consistent stellar population synthesis (SPS), mitigating potential overestimations of the stellar continuum caused by significant nebular emission.

    \item Present-day clump stellar masses were determined from the SPS results, as well as their luminosity- and mass-weighted mean ages and stellar metallicities. The best-fit contributions of the simple stellar populations (SSPs) were presented in age bins, revealing that the $t \leq 10\,$Myr component within the clumps corresponds to very metal-poor ($0.005 \lesssim Z_{\ast} / Z_{\odot} \lesssim 0.02$) populations, except for clump A, which shows solar metallicity in this age bin. However, as discussed in Appendix~\ref{appendix:SPS_caveats}, these solutions are highly degenerate, making it very difficult to determine the stellar metallicities accurately.
    Despite being identified from a continuum-subtracted H$\alpha$ image, the dominant stellar emission originates always from populations with $t >100\,$Myr. This occur likely because the clumps, located within the disk of \catname, are significantly smaller than the point spread function (PSF) size of MUSE observations, causing the integrated light of the clumps within PSF-sized apertures to include emission from the underlying disk.

    \item A potential consequence of the effect described above is an overestimation of the SPS-derived present-day clump stellar masses \citep[see][for example]{Ambachew2022}.
    Nevertheless, the raw clump stellar masses in \catname, estimated to lie in the range of $7.4 \lesssim \log(M/M_{\odot}) \lesssim 8.3$, are in fair agreement with raw and disk-subtracted clump populations within galaxies from the DYNAMO survey \citep{Green2014, Ambachew2022}, as well as with the stellar masses of the clump sample from $z \gtrsim 1$ lensed galaxies observed with JWST NIRSpec/NIRCam within the SMACS J0723.3–7323 cluster field  \citep{Claeyssens2023}.
    
    \item H$\alpha$ channel maps demonstrate the clumps are displaced relative to the nucleus, residing within the rotating ionized gas of \catname. L21 previously proposed the presence of a disk in \catname, based on the rotational pattern observed in the $v$(H$\alpha$) map. This suggests clump formation through violent disk instability (VDI). However, SPS results indicate that the youngest stellar populations within the clumps are more metal-poor than the average gas-phase metallicity ($Z_{\mathrm{gas}}$\,$\sim$\,0.3\,$Z_{\odot}$) observed across \catname.

    \item Infall of metal-poor gas as the trigger of clumpy SF in \catname was investigated by examination of $Z_{\mathrm{gas}}$ versus $\Sigma_{\mathrm{H}\alpha}$, which is expected to show an anti-correlation if SF is fuelled by metal-poor gas. Both strong-line calibrations tested show remarkably flat metallicity distributions instead. A mild yet systematic 0.1\,dex in $Z_{\mathrm{gas}}$ is found for $\log \Sigma_{\mathrm{H}\alpha} \gtrsim -15.7$ in one of the cases.
\end{itemize}

Flat metallicity distributions are common among dwarf irregular galaxies and can result from galactic-scale processes capable to efficiently mix metals in disk. While the lack of a clear $Z_{\mathrm{gas}}$ vs. $\Sigma_{\mathrm{H}\alpha}$ anti-correlation argues against infall of metal-poor gas as the primary trigger of clumpy star formation, it is possible that, as with the SPS results, the spectra contribution is dominated by the background disk, even at the bin resolution. Interestingly, the M13-derived values show a subtle ($\sim$0.1\,dex) anti-correlation arising at the high end of the $\Sigma_{\mathrm{H}\alpha}$ range. These data points correspond to single-spaxel resolution values with highest SFRs, where recent SF episodes fuelled by the infall of metal-poor gas could outshine the background light from the ionized extended component locally. 

SPS, on other hand, is more effective at disentangling the recently-formed clump population from the integrated light. The consistent contributions from young, metal-poor stellar populations within the clumps provide evidence supporting the ex situ scenario for clump formation. We note, however, that the solutions are highly degenerate, and the evidence for very metal-poor stellar populations in the clumps of \catname remains tentative, requiring stronger observational support for robust confirmation.

Unfortunately, the relatively large distance of \catname imposes spatial resolution limits that prevents precise constraints on the clump sizes. 
Hydrodynamical simulations represent a powerful tool to investigate whether \catname resulted from a past merger between low-mass progenitors or by ongoing tidal disruption, since pair collisions are one possible mechanism for producing flat metallicity gradients \citep[see the case of the extremely metal-poor dwarf irregular galaxy DDO 68,][]{Pascale2022}. Simulations can also help addressing how the forced alignment of the ionized gas in \catname relates to the formation of the kpc-sized clumps. For instance, could this alignment have triggered regions of instability within \catname? Is the timescale of clump formation connected to the approach of the interacting galaxies? And what is the fate of the clumps in \catname? Can they be disrupted and coalesce into the dwarf ISM, or will \catname be accreted before this occurs? While these questions are beyond the scope of this paper, hydrodynamical simulations are extremely useful for future studies aimed at addressing these and other aspects of this intriguing system.

\section*{Acknowledgements}
We thank the referee for constructive comments and suggestions that helped to improve the paper.
A.E.L. thanks Cid Fernandes and Angela Adamo for the fruitful discussions and also Natalia Vale Asari, Miriani Pastoriza and Charles Bonatto for the insightful feedback about this work during its earliest phases. A.E.L. acknowledges the support from \textit{Coordena\c{c}\~ao de Aperfei\c{c}oamento de Pessoal de N\'{i}vel Superior} (CAPES) in the scope of the Program CAPES-PrInt, process number 88887.837405/2023-00 and CAPES-PROEX fellowship, process number 88887.513351/2020-00.\,
A.C.S. acknowledges funding from the \textit{Conselho Nacional de Desenvolvimento Científico e Tecnológico} (CNPq) and the \textit{Funda\c{c}\~ao de amparo \`{a} pesquisa do Rio Grande do Sul} (FAPERGS) through grants  CNPq-314301/2021-6, FAPERGS/CAPES 19/2551-0000696-9.\,
R.R. thanks to CNPq (Proj. 311223/2020-6,  304927/2017-1 and  400352/2016-8), FAPERGS (Proj. 16/2551-0000251-7 and 19/1750-2), and CAPES (Proj. 0001).\,
E.J.J. acknowledges support from FONDECYT Iniciaci\'on en investigaci\'on 2020 Project 11200263 and the ANID BASAL project FB210003.\,
R.E.G.M. acknowledges support from the Brazilian agency CNPq through grants 406908/2018-4 and 307205/2021-5, and from \textit{Funda\c c\~ao de Apoio \`a Ci\^encia, Tecnologia e Inova\c c\~ao do Paran\'a} through grant 18.148.096-3 -- NAPI \textit{Fen\^omenos Extremos do Universo}.\,
R.A.R acknowledges the support from CNPq (Proj. 303450/2022-3, 403398/2023-1, \& 441722/2023-7), FAPERGS (Proj. 21/2551-0002018-0), and CAPES (Proj. 88887.894973/2023-00).\,
G.M.A. thanks CAPES for the project 88887.629085/2021-00, CAPES-PrInt program 88887.936546/2024-00, and the people of the Departamento de F\'{i}sica Te\'{o}rica y del Cosmos for receiving him very well in the University of Granada.\,
J.M.G. expresses gratitude to FCT for supporting years of research in Portugal, as well as to the Fundo Europeu de Desenvolvimento Regional 
(FEDER) through COMPETE2020–Programa Operacional Competitividade. J.M.G. also acknowledges the FCT-CAPES Transnational Cooperation Project. JMG had a DL 57/2016/CP1364/CT0003 contract and past fellowships (within the context of the FCT project UID/FIS/04434/2013) and POCI-01-0145-FEDER-007672. JMG was further funded by SFRH/BPD/66958/2009, supported by FCT and POPH/FSE (EC).\,
Besides the Python packages mentioned throughout the manuscript, this work made extensive use of \textsc{Astropy}\footnote{\label{foot:astropy_link}\url{http://www.astropy.org}}, a community-developed core Python package and an ecosystem of tools and resources for astronomy \citep{astropy:2022}, \textsc{uncertainties}
\footnote{\url{https://github.com/lmfit/uncertainties}}, and \textsc{SpectRes} \citep{Carnall2017}.

\section*{Data Availability}
All the data used in this work are public and are available at ESO Science Archive Facility under the program-ID of 099.B-0411(A) and at the Gemini Observatory Archive (GOA) under the program-ID GS-2021B-DD-103.


\bibliographystyle{mnras}
\bibliography{ref.bib}
\newpage
\appendix

\section{Stellar Population Synthesis: Interpretations and Caveats}
\label{appendix:SPS_caveats}

In the SF regions B, C, D, F, G and K, the SPS results shown in Fig.~\ref{fig:synth_results} suggest the presence of an ultra-low metallicity (ULM) $t < 10\,$Myr stellar population. Except for clump K, in all these regions this ULM subpopulation has light-weighted contributions above 0\% within the lower bounds provided by the 300 MCMC iterations.
The presence of such an intriguing subpopulation has a notable implication: given that the gas-phase metallicities lie in the range $0.2 \lesssim Z/Z_{\odot} \lesssim 0.3$ (see Table~\ref{tab:clump_properties}), any infalling pristine gas would have to form stars before mixing with the ISM.

In contrast, the nucleus and clump H show a $t < 10\,$Myr stellar population with $Z_{\ast} = 0.2\,Z_{\odot}$, similar to the measured gas-phase metallicity. Clump A, on the other hand, show solar metallicities for this age bin. Notably, clump B is located nearby clump A, with the PSF-sized circular apertures used to extract their spectra partially overlapping. Therefore, consistent stellar population properties would have been expected, even though they sample a physical region of approximately 1\,kpc in radius.

Looking to clarify this and the highly mixed subpopulations observed in the older age bins, we repeat the SPS analysis for all clumps shown in Fig.~\ref{fig:synth_results}, this time employing alternative SSP grids. To quantitatively compare these results with the fiducial run, we evaluate the fit quality using the $\chi^2$ and ADEV parameters, defined as follows \citep{Gomes2017}:

\begin{align}
    &\chi^2 = \sum\limits_{\lambda}^{} \Bigg[\frac{(f_{\lambda}^{\mathrm{syn}} - f_{\lambda}^{\mathrm{obs}})}{e_{\lambda}}\Bigg]^2 \\
    &\mathrm{ADEV} = \sum\limits_{\lambda}^{} \frac{\left| f_{\lambda}^{\mathrm{syn}} - f_{\lambda}^{\mathrm{obs}} \right|}{f_{\lambda}^{\mathrm{obs}}}
\end{align}

\noindent where $f_{\lambda}^{\mathrm{syn}}$ and $f_{\lambda}^{\mathrm{obs}}$ are the synthetic and observed spectra, respectively, and $e_{\lambda}$ is the error spectrum.
We tested four different SSP grid configurations: \textit{i)} Using $Z_{\star} = 0.02\,Z_{\odot}$ as the lowest metallicity;
\textit{ii)} Using $Z_{\star} = 0.2\,Z_{\odot}$ as the lowest metallicity;
\textit{iii)} Adopting solar metallicity as the highest metallicity;
\textit{iv)} Combining $Z_{\star} = 0.2\,Z_{\odot}$ as the lowest metallicity and solar metallicity as the highest in the grid.

\begin{table*}
\caption{Quantitative comparison of fit quality -- measured using $\chi^2$ and ADEV metrics (definitions in the text) -- variations with respect to the fiducial case (all stellar metallicities from \protect\cite{Bruzual2003} included). Values in bold indicate cases where the synthesis failed to reproduce the observed spectrum.}
\label{tab:SPS_stats}
\centering
\small
\begin{tabular}{l|cc|cc|cc|cc|cc}
\toprule
\toprule
\multirow{2}{*}{Clump ID} &
  \multicolumn{2}{c|}{Fiducial} &
  \multicolumn{2}{c|}{$Z_{\mathrm{min}} = 0.02\,Z_{\odot}$} &
  \multicolumn{2}{c|}{$Z_{\mathrm{min}} = 0.2\,Z_{\odot}$} &
  \multicolumn{2}{c|}{$Z_{\mathrm{max}} = Z_{\odot}$} &
  \multicolumn{2}{c}{$Z_{\mathrm{max}}, Z_{\mathrm{min}} = (Z_{\odot}, 0.2\,Z_{\odot})$} \\ \cline{2-11} 
 & \makebox[0pt][c]{$\chi^2$} & ADEV 
 & \makebox[0pt][c]{$\chi^2$} & ADEV 
 & \makebox[0pt][c]{$\chi^2$} & ADEV 
 & \makebox[0pt][c]{$\chi^2$} & ADEV 
 & \makebox[0pt][c]{$\chi^2$} & ADEV \\
\hline
Nucleus & 2.69 & 3.51 & 2.62 & 3.54 & 2.55 & 3.59 & 2.63 & 3.56 & 2.50 & 3.56 \\
A       & 1.48 & 9.09 & 1.46 & 9.02 & 1.44 & 9.03 & 1.45 & 8.97 & 1.41 & 9.08 \\
B       & 1.58 & 6.43 & 1.55 & 6.43 & 1.52 & 6.58 & 1.54 & 6.49 & 1.48 & 6.56 \\
C       & 2.43 & 6.49 & 2.37 & 6.14 & 2.32 & 6.24 & 2.37 & 6.22 & 2.27 & 6.16 \\
D       & 2.24 & 3.40 & 2.21 & 3.35 & 2.15 & 3.47 & 2.20 & 3.40 & 2.11 & 3.44 \\
F       & 2.30 & 4.53 & 2.25 & 4.49 & 2.20 & 4.51 & 2.25 & 4.51 & 2.16 & 4.49 \\
G       & 2.22 & 4.81 & 2.18 & 4.84 & 2.20 & 4.82 & 2.17 & 4.82 & 2.11 & 4.82 \\
H       & 2.85 & 9.10 & \pmb{3.57} & 8.35 & \pmb{3.43} & 8.37 & 2.70 & 8.72 & \pmb{3.52} & 9.26 \\
K       & 3.16 & 2.37 & 3.10 & 2.37 & 3.04 & 2.39 & 3.10 & 2.36 & 2.98 & 2.40 \\
\bottomrule
\bottomrule
\end{tabular}

\end{table*}

The results are summarized in Table~\ref{tab:SPS_stats}. Except for clump H, the fit quality remains remarkably similar across all configurations, regardless of whether $\chi^2$ or ADEV is used as the metric. As shown in Fig.~\ref{fig:synth_results}, the inclusion of ULM models in the grid leads to non-negligible contributions. However,
Table \ref{tab:SPS_stats} also reveals that the subpopulation solutions are highly degenerate. While the different SSP grids have little to no effect on the mean stellar ages reported in Table~\ref{tab:clump_properties}, the same is evidently not true for the mean stellar metallicities. Therefore, any interpretations drawn from comparisons between the disc nebular metallicity and the clump stellar metallicities require caution. Finally, clump H represents an interesting exception. The boldface values in Table~\ref{tab:SPS_stats} highlight the cases in which the synthesis performed poorly in reproducing the observed stellar emission. While the synthesis results for clump H remain nearly unchanged when the supersolar metallicity is removed from the grid, without the inclusion of ULM models in the grid, the observed stellar emission in this clump cannot be satisfactorily reproduced. We emphasise, however, that this is not sufficient to claim the existence of ULM populations in \catname, although it justifies the use of the complete model grid rather than a simplified one restricted to sub-solar, solar, and super-solar metallicities.

\section{High-resolution NIR imaging with GSAOI}
\label{appendix:GSAOI}

\begin{figure*}
    \centering
    \includegraphics[]{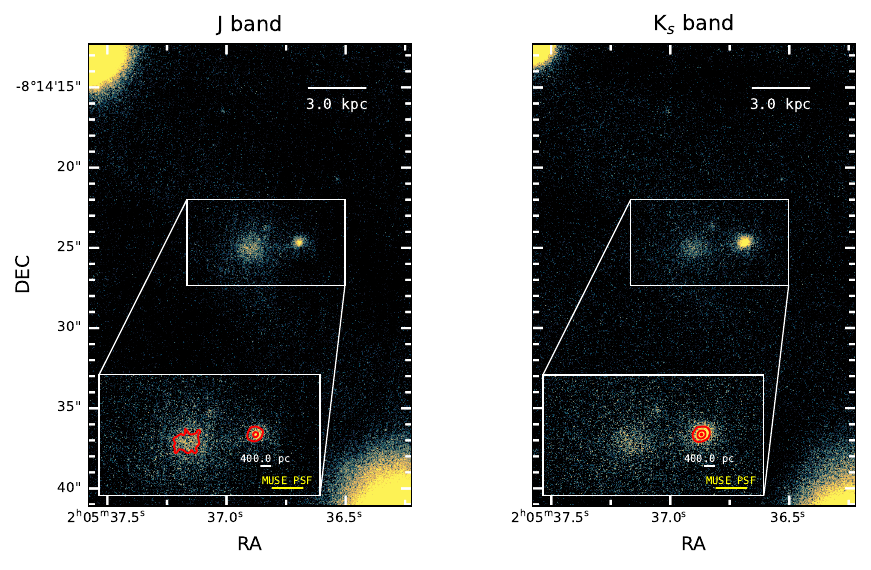}
    \caption{Zoom-in of GSAOI FoV on \catname for J and \Ks bands. The inset plots on both frames present a zoom-in on the region corresponding to the emission peak in the optical continuum and the emission peak in H$\alpha$ (clump A). The red contours represent the levels of the median background + (2, 3)$\times \sigma$ for J band and the median background + (2, 3, 4)$\times \sigma$ for \Ks band. The white scale bar indicates the approximate diameter of 400\,pc that is estimated for clump A, whereas the yellow scale bar represent the PSF size of our MUSE observations ($\sim$~1.1\,kpc).}
    \label{fig:GSAOI}
\end{figure*}

\subsection{Observation and data reduction}
Observations of \catname were also carried with the high-resolution AO-assisted imaging in the NIR from the Gemini South Adaptive Optics Imager
\citep[GSAOI,][]{Mcgregor2004, Carrasco2012} with the Gemini Multi-Conjugate Adaptive Optics System (GeMS), a sodium-based multi-Laser Guide adaptive optics system mounted at the Gemini-South telescope in Cerro Pachón \citep{rigaut2014, neichel2014}. GSAOI is a NIR camera located at the $f$/32 output focus of Canopus and delivers images near diffraction limit in the wavelength range 0.95\,--\,2.5\,$\mu$m. The instrument covers a FoV of approximately 85\,$\times$\,85\,arcsec$^{2}$ with a spatial scale of 0.02\arcsec per pixel. The GSAOI detector is formed by four Rockwell HAWAII-2RG arrays mounted in a 2\,$\times$\,2 mosaic that creates a 4080\,$\times$\,4080 pixel focal plane. The gap between the arrays is $\sim$~2.5\,mm, which corresponds to $\sim$~2.4\arcsec on the sky. The target was observed (Program-ID: GS-2021B-DD-103) using the GSAOI filters J (1.250\,$\mu$m) and \Ks\,(2.150\,$\mu$m) in 20 and 21 exposures, respectively, with an exposure time of 120\,s each, resulting in total observing times of 40\,min and 42\,min. A dithering pattern was adopted around the target and separate sky exposures were not requested. The seeing delivered is 0.18\arcsec and 0.15\arcsec in bands J and \Ks, respectively.

The data reduction was performed using \textsc{theli} software package \citep{Erben2005, schirmer2013}. Initially, the Header Data Unit (HDU) of the 41 raw images is reformatted to split each exposure into separate files, one for each chip, to allow parallelisation. Detector-level effects such as non-linearity and cross-talk are corrected and the gain of each frame is adjusted. A set of 15 flat field observations with the dome lamps on was taken for each band. Additionally, a set of 15 flat exposures were taken with the dome lamps off for \Ks band with an exposure time of 15\,s. Those are used to create a master flat field which is applied to each array. To model the background we adopt the dynamic mode in \textsc{theli}, given that for image sequences in the NIR exceeding 30\,min temporal variations of the background due to multiplicative and additive effects (e.g. fringing, moonlight and atmospheric airglow) become non-negligible \citep{Schirmer2015}. The dynamic mode groups the exposures in 4 groups of 5 distinct exposures to mitigate background variations. After a standard collapse correction, the reduced sky-subtracted exposures of each band are resampled and co-added using \textsc{swarp} \citep{Bertin2010}.

\subsection{Sizes of SF regions}
In Fig.~\ref{fig:GSAOI} we present a zoom-in of GSAOI FoV on the same region presented in the right panels of Fig.~\ref{fig:MUSE_FoV}, covering \catname. Most of the dwarf emission is below the background level, given that in the outermost regions of \catname the emission is almost entirely originated from the ionised gas. However, the region corresponding to clump A is clearly observed, and it is also possible to see a faint and more extended emission in the region corresponding to the peak of the continuum emission. In the zoom-in insets, we present contours that represent the background median + 2,3$\sigma$ in the J band and the background median + 2,3,4$\sigma$ in the \Ks band. With the aid of the AO-assisted high-resolution image delivered by GSAOI it is possible to see that the diameter of clump A is $\approx$~400\,pc, smaller than the $\sim$~1\,kpc that we derive based on MUSE observations.

\bsp	
\label{lastpage}
\end{document}